\documentclass{emulateapj}
\usepackage{graphicx,amsmath}
\usepackage{color}
\usepackage{booktabs}
\usepackage{times}
\usepackage{float}
\usepackage{ulem}
\usepackage{multirow}

\def\be{\begin{equation}}
\def\ee{\end{equation}}
\def\bea{\begin{eqnarray}}
\def\eea{\end{eqnarray}}

\shorttitle{Variations of Broad Emission Lines from QSOs with Periodicity}
\shortauthors{Ji et al.}

\begin{document}

\title{
Variation of Broad Emission Lines from QSOs with Optical/UV Periodicity to Test 
the Interpretation of Supermassive Binary Black Holes}

\author{Xiang Ji\altaffilmark{1,2},
        Youjun Lu\altaffilmark{1,2}, 
        Junqiang Ge\altaffilmark{1},
        Changshuo Yan\altaffilmark{1,2},
        and Zihao Song\altaffilmark{1,2}}

 \affil{  
     $^{1}$ CAS Key Laboratory for Computational Astrophysics, National Astronomical Observatories, Chinese Academy of Sciences, 20A Datun Road, Beijing 100101, China; luyj@nao.cas.cn\\
     $^{2}$ School of Astronomy and Space Science, University of Chinese Academy of Sciences, No. 19A Yuquan Road, Beijing, 100049, China\\
}

\begin{abstract} 
Periodic quasars have been suggested to host supermassive binary black holes (BBHs) in their centers, and their optical/UV periodicities are interpreted as caused by either the Doppler-boosting (DB) effect of continuum emission from the disk around the secondary black hole (BH) or intrinsic accretion rate variation. However, no other definitive evidence has been found to confirm such a BBH interpretation(s). In this paper, we investigate the responses of broad emission lines (BELs) to the continuum variations for these quasars under two BBH scenarios, and check whether they can be distinguished from each other and from that of a single BH system. We assume a simple circumbinary broad-line region (BLR) model, compatible with BLR size estimates, with a standard $\Gamma$ distribution of BLR clouds. We find that BELs may change significantly and periodically under the BBH scenarios due to (1) the position variation of the secondary BH and (2) the DB effect, if significant, and/or intrinsic variation, which is significantly different from the case of a single BH system. For the two BBH scenarios, the responses of BELs to (apparent) continuum variations, caused by the DB effect or intrinsic rate variation, are also significantly different from each other, mainly because the DB effect has a preferred direction along the direction of motion of the secondary BH, while that due to intrinsic variation does not. Such differences in the responses of BELs from different scenarios may offer a robust way to distinguish different interpretations of periodic quasars and to identify BBHs, if any, in these systems.    
\end{abstract}

\section{Introduction}
\label{sec:intro}

Supermassive binary black holes (BBHs) are natural products of galaxy mergers \citep[e.g.,][]{1980Natur.287..307B, Yu02} and are expected to exist in the centers of many galaxies \citep[e.g.,][]{2003ApJ...582..559V, CYL20}. Observations have revealed a number of supermassive black hole (BH) pairs with separations on kiloparsec scales \citep[e.g.,][]{2010ApJ...715L..30L, 2011ApJ...737..101L, 2018ApJ...862...29L,  2009ApJ...702L..82C, 2013ApJ...777...64C, 2015ApJ...806..219C, 2011ApJ...735...48S, 2012ApJS..201...31G, 2012ApJ...746L..22K}, whose galactic nuclei are still in the process of merging \citep[e.g.,][]{2011ApJ...738...92Y, 2019RAA....19..177Y, 2019SCPMA..6229511Y} and will ultimately evolve to become close-bound BBHs with separation smaller than subparsec scales. However, observational evidence for subparsec-scale BBHs are still elusive despite efforts over several decades in searching for them. This is not only due to the sizes of these systems, which are too small to be spatially resolved \citep[e.g.,][]{Yu02, 2011MNRAS.410.2113B}, but also due to the lack of distinct and unique signatures of these systems that can be used to distinguish them from alternatives \cite[e.g.,][]{2012NewAR..56...74P, 2004ApJ...613L..33G}.

In the past several decades, nevertheless, more than 100 BBH candidates have been found by using some characteristic features as indicators for the existence of BBHs in active galactic nuclei \citep[AGNs; see][for an overview]{2020RAA....20..160W}. Such features include the periodic variation of the continuum or broad emission lines \citep[BELs; e.g.,][]{1988ApJ...325..628S, Graham15, 2015MNRAS.453.1562G, 2016MNRAS.463.2145C, 2018MNRAS.476.4617C, 2016ApJ...822....4L,  2019ApJS..241...33L}, the optical/UV deficit in the continuum spectrum of a quasar \citep[e.g.,][]{2015ApJ...809..117Y, Zheng16}, and the double-peaked or asymmetric shapes of BELs \citep[e.g.,][]{BL09, Tsalmantza11, Eracleous12, 2012ApJ...759..118B, 2012NewAR..56...74P, 2013ApJ...777...44J, 2013ApJ...775...49S, 2011ApJ...735...48S, 2014ApJ...789..140L, 2016ApJ...822....4L,  2019MNRAS.482.3288G}. Recently, even the changing look of an AGN was attributed to the dynamical effect of BBHs \citep[see][]{2020A&A...643L...9W}.

Among the BBH candidates, one of the currently hotly debated ones is PG 1302-102, with periodic optical/UV/radio variations on a period of $\sim 5$\,yr \citep[see][but \citealt{2018ApJ...859L..12L}]{Graham15, 2018A&A...615A.123Q, 2020MNRAS.496.1683X}. Its sinusoid-like light curves can be well explained by the relativistic Doppler-boosted/-weakened emission mainly from the accretion disk around the secondary component of a BBH system \citep{2015Natur.525..351D}. 

However, there may be some deviations of the observed light curves (or line profiles) from those predicted by the Doppler-boosting interpretation as pointed out by \citet{2018A&A...615A.123Q} and  \citet[or \citealt{2020MNRAS.491.4023S, 2021A&A...645A..15S}]{2019ApJ...871...32K}. Such sinusoidal light curves may also be directly due to the periodical variation of the accretion rate modulated by the BBH orbital motion as suggested by numerical simulations \citep[e.g.,][]{2008ApJ...682.1134H, 2008ApJ...672...83M,  2014MNRAS.439.3476R, 2015MNRAS.447L..80F,  2017ApJ...838...42B, 2018ApJ...853L..17B}. \citet{2018ApJ...859L..12L} even doubted the periodicity in the light curves because of some subsequent observations, which pose a significant challenge to the BBH interpretation of PG 1302-102. Furthermore, it has not been ruled out that the ``periodicity'' of optical/UV variation may also occur in a single BH accretion system for some unknown reasons. For example, the ``periodicity'' found in some of the quasars reported by \citet{2015MNRAS.453.1562G} and \citet{2016MNRAS.463.2145C} may be a false few-cycle one due to stochastic variability over a wide range of timescales \citep[][]{2016MNRAS.461.3145V}.

To identify/falsify BBH candidates, including those suggested by the periodic light curves of quasars, it is important to find multiple BBH signatures or multiple lines of evidence. One way might be combining the BELs and their variation with the periodic light curves. Therefore, it is
intriguing to theoretically investigate BEL (variation) features for hypothetical BBH systems with periodic continuum variations (e.g., similar to PG 1302-102). If the variation patterns of BELs from BBH systems are distinct from those of single BH systems with a similar continuum variation, they can be used as unique signatures to confirm the BBH systems via the reverberation mapping (RM) technique \citep[e.g.,][]{1972ApJ...171..467B, 1982ApJ...255..419B, 1993PASP..105..247P, 2004AN....325..248P, Bentz2016}.

There have been a number of studies to investigate the detailed variation pattern of BELs from subparsec BBH systems \citep[e.g.,][]{ 2010ApJ...725..249S, 2018ApJ...862..171W, 2020ApJS..247....3S, 2020A&A...635A...1K}; these mainly focused on the line profiles and their variations
for those BBH systems with two separate BLRs associated with each of the two BH components. Recently, \citet{2018ApJ...862..171W} and \citet{2020ApJS..247....3S} have shown the distinctive structures of BBH systems in the two-dimensional transfer functions (2DTFs) and demonstrated the possibility of using the RM technique to reveal BBH systems \citep[see also][]{2020A&A...635A...1K}. We note here that the BBH candidates suggested by periodicity continuum variations \citep{2015MNRAS.453.1562G, 2016MNRAS.463.2145C} may have BLRs with sizes substantially larger than the inferred BBH semimajor axes \citep[e.g.,][]{2020MNRAS.491.4023S, 2021A&A...645A..15S}, i.e., circumbinary BLRs, which are in a different category from those cases with two separate BLRs in the above-mentioned studies. In this paper, we investigate the response of BEL profiles to the continuum changes for BBH/BH systems with periodic variation, as either a result of the DB scenario or the intrinsic variation of the central source, explore whether it is possible to distinguish different scenarios for the periodic quasars with similar continuum variations, and confirm the BBH scenario(s) by using the variations of its BELs.

This paper is organized as follows. In Section~\ref{sec:model}, we introduce a simple model for BBH systems with circumbinary BLRs, derive a framework for the response of BELs to (periodic) continuum variations under both the DB scenario for BBH systems and the intrinsic variation scenario for both BBH and single BH systems. Note that we do not consider other variation mechanisms, such as the damped random walk \citep[e.g.][]{2016MNRAS.461.3145V}, in the present paper for simplicity. We present our model settings in Section~\ref{sec:modelset}. We further analyze the differences in the responses of BEL profiles to periodic continuum variation resulting from different scenarios, and show that such differences can be used as unique signatures to identify/confirm BBH systems using multiple spectroscopic observations in Section~\ref{sec:results}. 
Discussions are given in Section~\ref{sec:discussion}. Conclusions are summarized in Section~\ref{sec:conclusion}.  

\section{Simple Models for Broad Line Emission}
\label{sec:model}

Suppose that the observed optical/UV continuum from a quasar varies periodically with sinusoid-like light curves, such as those of PG 1302-102 \citep[][]{Graham15}. The sinusoid-like light curves can be interpreted to be due to the relativistic DB/weakening of (nonvarying) continuum emission from a BBH system as suggested by \citet[][hereafter denoted as the BBH-DB scenario]{2015Natur.525..351D}. Alternatively, the periodic variation of the accretion rate(s) modulated by the orbital motion of a BBH system, as suggested by theoretical studies, may contribute significantly to or even be dominant in the observed continuum variation \citep[e.g.,][hereafter denoted the BBH-IntDB scenario as the DB effect may also contribute]{2008ApJ...682.1134H, 2008ApJ...672...83M,  2014MNRAS.439.3476R, 2015MNRAS.447L..80F,  2017ApJ...838...42B, 2018ApJ...853L..17B}. Furthermore, a QSO with a single BH in its center may also have a periodic optical/UV variation due to instability induced accretion rate variation, or a few-cycle false ``periodicity'' due to stochastic variations (hereafter denoted as the BH-Int scenario), or at least, such a possibility has not been ruled out yet. To reveal the nature of those quasars with optical/UV periodicity, it is important to distinguish these three different scenarios.
One may note that the intrinsic continuum emission from the source should be quite different in different scenarios, which may lead to different variations of the BELs. In order to estimate this, it is necessary to first know the geometric configuration of the BLR and the cloud distribution around the BBH, and then BEL profiles and their variations can be obtained in the circumstances of different continuum variation scenarios. In this section, we introduce a framework for considering the response of BELs to periodic continuum variation under different scenarios, as detailed below.

\subsection{BBH-Circumbinary BLR configuration}

\begin{figure} 
\centering
\includegraphics[width=0.5\textwidth]{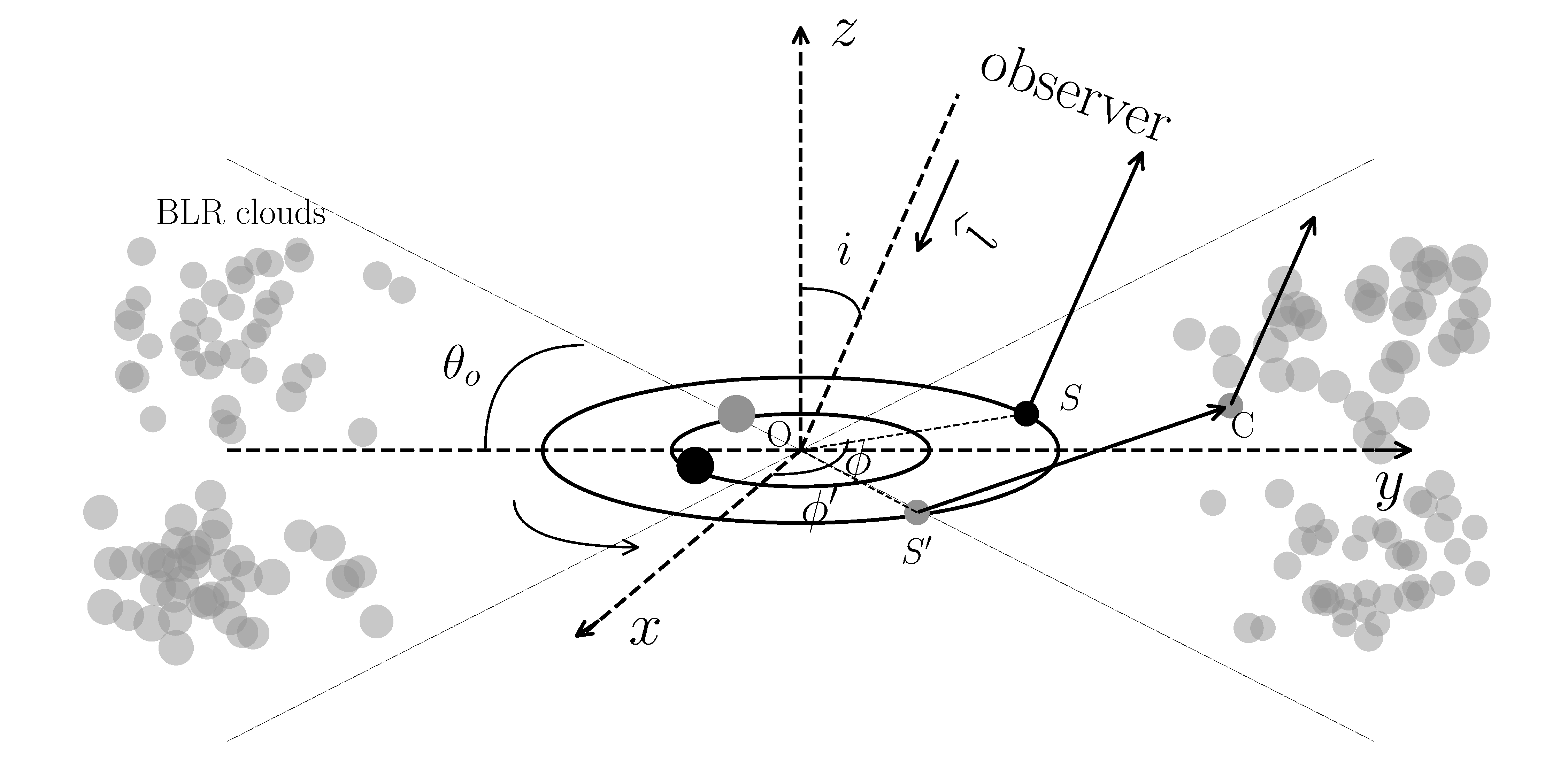}
\caption{Sketch diagram for the structure of a BBH system with a circumbinary BLR. Suppose that the two primary (large solid circles) and secondary components (small solid circles) rotate counterclockwise on circular orbits in the $xy$ plane around the mass center. The black and gray solid circles represent the positions of the two components at two observation times. Gray solid points represent BLR clouds rotating around the BBH. C represents an arbitrary cloud in the BLR, $i$ denotes the viewing angle of the BBH orbital plane, i.e., the angle between  the direction of the line of sight ($\hat{l}=(0,-\sin i, -\cos i)$) and the $z$-axis, and $\theta_{\rm o}$ represents the opening angle of the BLR. At a specific observation time $t_{\rm obs}$, the observer receives continuum photons emitted from the disk around the secondary BH at $S$ and the orbital phase is $\phi$. At the same time, the observer also receives BEL photons reemitted by cloud C, which were induced by the illumination of continuum flux emitted from the disk around the secondary BH at $S^{\prime}$ with an orbital phase of $\phi'$.
}
\label{fig:f1}
\end{figure} 

Figure~\ref{fig:f1} shows a sketch diagram for a BBH system with a circumbinary BLR. Suppose at the moment of observation $t_{\rm obs}$, a distant observer receives the optical/UV continuum emission mainly from the disk associated with the secondary BH when it is at the position $S$, considering the effect of time delay due to the BH motion. On the other hand, for a specific cloud C$_i$ in the BLR, it receives the ionizing photons emitted from the disk when the BH is at the position $S^{\prime}$, and this induces the emission of line photons instantly (ignoring the short recombination time), which are finally received by the observer at the same moment of observation $t_{\rm obs}$. 
For the continuum emission, the time in the observer's frame ($t_{\rm obs}$) can be related to the intrinsic time ($t_{\rm in}$) in the BBH mass center rest frame (hereafter BBH frame) considering of the motion of the secondary BH and cosmological time dilation, i.e., 
\begin{eqnarray}
t_{\rm obs}-t_{\rm obs,0} & = & \left[(t_{\rm in} -t_{\rm in,0})+ \frac{\overrightarrow{\rm \boldsymbol{OS} }\cdot {\rm \boldsymbol{\hat{l}} }}{c}\right](1+z),  \\
t_{\rm in} -t_{\rm in,0} & = &  T_{\rm orb} \phi/2\pi, \\
\overrightarrow{\rm \boldsymbol{OS} }\cdot {\rm \boldsymbol{\hat{l}} } & = &
-a_{\rm BBH} \sin i \sin \phi /(1+q),
\end{eqnarray}
where $t_{\rm obs,0}$ is the starting time for the observation, $t_{\rm in}$ and $t_{\rm in,0}$ are the times in the BBH frame corresponding to $t_{\rm obs}$ and $t_{\rm obs,0}$, respectively, $a_{\rm BBH}$ is the semimajor axis of the BBH system at redshift $z$, $q$ is the mass ratio, $c$ is the speed of light, $\phi_0$ is the phase for the secondary BH at $t_{\rm obs,0}$, and $\hat{\boldsymbol{l}}$ is the unit vector along the line of sight. For convenience, we set $t_{\rm obs,0}=0$, $t_{\rm in,0}=0$, and $\phi_0=0^{\circ}$, i.e., the secondary BH is located at the positive $x$-axis, when $t_{\rm obs,0}=0$.
As for BEL line photons emitted from cloud C, located at $(x_{0},y_{0},z_{0})$, the time in the observer's frame ($t_{\rm obs}$) can be related to the intrinsic time ($t_{\rm in}^{'}$) in the BBH frame considering the motion of the secondary BH, i.e., 
\begin{eqnarray}
& t_{\rm obs}-t_{\rm obs,0}  =  \left[ (t_{\rm in}^{'} -t_{\rm in,0}) + \frac{\overrightarrow{\rm \boldsymbol{OC} }\cdot \rm {\rm \boldsymbol{\hat{l}} }+ |\overrightarrow{\boldsymbol{S^{\prime}C} }|}{c} \right] (1+z), \nonumber \\ \\
& t_{\rm in}^{'} -t_{\rm in,0}  = T_{\rm orb} \phi^{\prime}/2\pi, \\
& \overrightarrow{\rm \boldsymbol{OC} }\cdot \rm {\rm \boldsymbol{\hat{l}} }  =  -(y_{0} \sin i +z_{0} \cos i), \\
& |\overrightarrow{\boldsymbol{S^{\prime}C} }|  = 
\left[\left(\frac{a_{\rm BBH}\cos\phi^{\prime}}{1+q}  -x_{0}\right)^{2} + \left(\frac{a_{\rm BBH}\sin \phi^{\prime} }{1+q} -y_{0} \right)^{2} +z_{0}^{2} \right]^{\frac{1}{2}},
\nonumber \\
\end{eqnarray}
where $t_{\rm in}^{'}$ is the time in the BBH frame when the secondary BH is located at $S^{\prime}$. The location of the cloud $(x_0, y_0, z_0)$ also changes with time due to its orbital motion. So, for a specific moment of observation $t_{\rm obs}$, we can derive both $\phi$ and $\phi^{\prime}$, and hence the emission time difference between the continuum received by the observer and that received by the cloud is 
\begin{eqnarray}
\tau = t_{\rm in} -t_{\rm in}^{'}
=\left( \overrightarrow{\rm \boldsymbol{OC} }\cdot \hat{\rm \boldsymbol{l} } +|\overrightarrow{\boldsymbol{S^{\prime}C} }| -
\overrightarrow{\rm \boldsymbol{OS} }\cdot \hat{\rm \boldsymbol{l} }\right) /c.
\end{eqnarray}
The above equations, derived for a BBH system with a circumbinary BLR, can be simply reduced to the general case of a BLR rotating around a single BH by setting $a_{\rm BBH}=0$, in which $\phi$ ($=\phi'$) represents the phase of the periodic variation of the continuum. 

\subsection{Continuum Variation}
\label{subsec:cont}

We explicitly consider three scenarios for the observed periodic variations of the optical/UV continuum found in some QSOs \citep[][]{Graham15, 2015MNRAS.453.1562G, 2016MNRAS.463.2145C}, i.e., (1) the BBH-DB scenario \citep[][]{2015Natur.525..351D}; (2) the BBH-IntDB scenario \citep[][]{2008ApJ...682.1134H, 2008ApJ...672...83M, 2014MNRAS.439.3476R, 2015MNRAS.447L..80F, 2017ApJ...838...42B, 2018ApJ...853L..17B}; and (3) the BH-Int scenario. 

In the BBH scenario, we assume that the intrinsic continuum flux emitted from the disk around the secondary BH can be described by a power law $F_{\nu, \rm e}(\nu; t) \propto \nu^{\alpha}$. This continuum may or may not vary with time. If the rotation velocity of the secondary BH is sufficiently high, then the continuum received by a distant observer can be Doppler boosted/weakened and vary periodically due to the modulation of the orbital motion. Thus, the observed flux at a given frequency $\nu_{\rm o}$ is
\begin{eqnarray}
F_{\nu,\rm o}(\nu_{\rm o}; t_{\rm obs}) & \propto &  D_{\rm sec}^3 F_{\nu,\rm e} \left(\frac{\nu_{\rm o}}{(1+z)D_{\rm sec}}; t_{\rm in} \right) \nonumber \\
& \propto & D_{\rm sec}^{3-\alpha} F_{\nu,\rm e}(\nu_{\rm e}; t_{\rm in}).
\end{eqnarray}
Here the relativistic Doppler factor, $D_{\rm sec}$, is determined by the relative motion ($\vec{v}_{\rm sec}$) of the secondary BH relative to the distant observer, i.e.,
\begin{equation}
D_{\rm sec} = \left[ \gamma_{\rm sec} \left(1-\beta \cos \phi \sin i \right)\right]^{-1},
\end{equation}
and $\beta= |v_{\rm sec}|/c$, the Lorentz factor $\gamma_{\rm sec} = 1/\sqrt{1-\beta^2}$, $\phi$ is the orbital phase, and $i$ is the inclination angle. When $\phi=0^\circ$ or $180^\circ$, $D_{\rm sec}$ is at its maximum or minimum value. Note that in the above equation, we ignore the constant factor due to cosmological redshift and $k$ correction.
If the intrinsic variation is negligible compared with that due to the DB, i.e., $F_{\nu,\rm e}(\nu_{\rm e}, t) \sim \textrm{const}$, then it is the BBH-DB scenario. In this scenario, the redshift-corrected observed period for the continuum variation is the same as the BBH orbital period. 

If the intrinsic variation is also periodic and significant or even dominant compared with that due to the DB effect depending on the BBH rotation velocity and viewing angle, then it is the BBH-IntDB scenario, where the $D_{\rm sec}^{3-\alpha}$ term can be important or negligible. In this scenario, the redshift-corrected observed period can be different from the BBH orbital period (e.g., half of the orbital period \citealt{2008ApJ...682.1134H}). Nevertheless, we assume it is the same as the BBH orbital period and the phase is also coincident with the DB effect for simplicity. In this scenario, the amplitude of the periodic variation due to DB ($\mathcal{A}_{\rm DB}$) is roughly given by 
\begin{eqnarray}
\log \mathcal{A}_{\rm DB} & \sim & \frac{1}{2} \left[ \log \left| D^{3-\alpha}_{\rm sec,max}-1\right| + \log \left| D^{3-\alpha}_{\rm sec,min} -1 \right|\right]. \nonumber \\ 
\label{eq: DBamp}
\end{eqnarray}
Note that $D_{\rm sec,max}$ and $D_{\rm sec,min}$ are the maximum and minimum DB factors at $\phi=0^\circ$ and $180^\circ$, respectively. The oscillation due to DB deviates from a sinusoidal curve and can be asymmetric because $D_{\rm sec,max}$ may be substantially larger than $ 1/D_{\rm sec,min}$ when the relative velocity $|\vec{v}_{\rm sec}|$ is close to the speed of light, though $|\vec{v}_{\rm sec}|$ is substantially smaller than the speed of light in most cases. We denote the amplitude of the intrinsic periodic variation as $\mathcal{A}_{\rm Int}$ and assume the phase at the maximum flux of this intrinsic variation is the same as that of the DB-effect-induced variation at $\phi=0^\circ$, for simplicity. If the amplitude of the observed continuum variation is $\mathcal{A}$, for a given BBH system as illustrated in Figure~\ref{fig:f1}, the contribution fraction from the intrinsic variation should be 
\begin{equation}
\mathcal{A}_{\rm Int} = \frac{\mathcal{A}+1}{\mathcal{A}_{\rm DB}+1} -1.
\label{eq:Intamp}
\end{equation}

For the third case, i.e., BH-Int, the continuum variation is due to the intrinsic variation, and it is irrelevant to any DB effect, i.e., $D_{\rm sec} \equiv 1$, therefore,
\begin{equation}
F_{\nu,\rm o}(\nu_{\rm o}; t_{\rm obs}) \propto F_{\nu, \rm e}(\nu_{\rm e}; t_{\rm in}), 
\end{equation}
where $t_{\rm in}= t_{\rm obs}/(1+z) + \textrm{const}$.
In this work, we set the spectral index $\alpha=-2$ \citep[][]{2015Natur.525..351D} for simplicity. 

\subsection{Structure and Kinematics of the Broad-line Region }
\label{subsec:BLRstructure}

In this subsection, we introduce a simple model for the BLR and the distribution of clouds in it. We assume the circumbinary BLR is flattened with an opening angle of $\theta_{\rm o}$ and the middle plane of the BLR is the same as the BBH orbital plane\footnote{In principle, the middle plane of the flattened BLR could be offset from the BBH orbital plane. We defer the detailed discussion of those cases with BLR offset from the BBH orbital plane to a separate work.}. The viewing angle of a distant observer is $i$, which is defined as the angle between the normal of BLR middle plane and the line of sight. See Figure~\ref{fig:f1} for illustration.

According to the studies of BLRs for QSOs (presumably most having a single central BH; \citealt[e.g.,][]{2015ARA&A..53..365N}, the inner radius of the BLR is larger than the outer radius of the accretion disk, normally assumed to cut at the self-gravity radius, and the maximum radius of the BLR cannot be larger than the sublimation radius. Therefore, we set the BLR within these two radii. We assume that the BLR is not expanding or shrinking because the dynamical timescale for the BLR clouds is usually much longer than the observation periods considered in the present paper (or for typical RM studies). We further assume that the radial emissivity distribution of the clouds in BLR can be described by a shifted $\Gamma$-distribution function, as in \citet[]{2014MNRAS.445.3055P}, i.e.,
\begin{eqnarray}
R_{\rm ga} = R_{\rm S} + F R_{\rm BLR} +g(1-F) \beta^{2} R_{\rm BLR},
\label{eq:Rga}
\end{eqnarray}
where $R_{\rm S} $ is the Schwarzschild radius, $R_{\rm BLR}$ is the mean distance of BLR clouds from the central source, $F = R_{\rm min}/R_{\rm BLR}$ is the fractional inner radius, $\beta$ is the shape parameter, and $g$  is drawn randomly from a $\Gamma$-distribution. The $\Gamma$-distribution for a positive variable $x$ is given in the form as:
\begin{equation}
P(x|\alpha,\theta) = x^{\alpha-1} \exp (-x/\theta)/\theta^\alpha,
\end{equation}
where $\theta$ is a scale parameter and $\alpha$ is the shape parameter. In this paper, we set $\beta=0.9$, $F=0.5$, $\theta=1$, and $\alpha = 1/\beta^{2}$ \citep[][]{2014MNRAS.445.3055P} for simplicity. 

For the system with a single central BH, the BLR clouds are illuminated by the ionizing photons from the central source, and the mean $R_{\rm BLR}$ can be estimated by using the empirical relation $R_{\rm BLR} \propto L^{1/2}$ \citep[][]{2000ApJ...533..631K, 2007ApJ...659..997K, 2002MNRAS.337..109M, 2004ApJ...613..682P} as
\begin{equation}
R_{\rm BLR}  \approx 2.2 \times 10^{-2}   {\rm pc} \left (\frac{\lambda_{\rm Edd}}{0.1} 
\right) ^{1/2} \left (\frac{M_{\bullet}}{10^{8} M_{\odot}} \right)^{1/2},
\label{eq:RBLR}
\end{equation}
where $M_{\bullet}$ is the BH mass and $\lambda_{\rm Edd}$ the Eddington ratio. 
We assume that each BLR cloud at a given $R_{\rm ga}$ is on a circular orbit, and the orbital velocity is determined by its radius,  i.e., $v_{\rm ga} = (GM_{\rm tot} / R_{\rm ga})^{1/2}$. The distribution of clouds in the BLR may be flattened but not spherically distributed as illustrated in Figure~\ref{fig:f1}. We assume that the angular distribution of the BLR clouds can be described by 
\begin{equation}         
\theta = \arccos \left[ \cos \theta_{\rm o} +(1-\cos \theta_{\rm o})   U^{\gamma} \right],
\label{eq:theta}
\end{equation}
where $\theta$ represents the angle between the orbital angular momentum of a cloud and the normal to the BLR middle plane, $U$ is a random number, and $U^{\gamma}$ is introduced to describe the concentration of  BLR clouds \citep[see][for details]{2014MNRAS.445.3055P}. In this paper, for simplicity, we set $U^{\gamma}$ to be a random number in the range from $0$ to $1$, which means that BLR cloud momenta are uniformly distributed but constrained by $\theta \leq \theta_{\rm o}$. We assume that all BLR clouds rotate around the central BH roughly in the same direction, either clockwise or counter-clockwise. This assumption is compatible with the case of 3C 273, the BLR of which  is resolved by \citet[]{2018Natur.563..657G}.

For the case of a BBH system with circumbinary BLR, we assume that the inner boundary of the BLR is about $2 a_{\rm BBH}$, at roughly the outer boundary of the gap opened by the secondary BH \citep[e.g.,][]{2014ApJ...783..134F, 2013MNRAS.436.2997D}. We adopt the shifted $\Gamma$ function for the radial emissivity distribution of the clouds (test particles) in the BLR rotating around the BBH mass center, similar to that for the single BH system. We assume that BLR clouds corotate with the BBH if not otherwise stated. We also consider the cases of BLR clouds counter-rotating around the BBHs, and investigate how the line variation pattern depends on the rotation direction of BLR clouds with respect to the BBH rotation direction in Section~\ref{sec:results}.

In the BBH cases, the orbital motions of BLR clouds (test particles) may deviate from those in the cases for single BH systems due to the dynamical interactions of these particles with the BBH. To account for this effect, we randomly generate the initial position and velocity of each particle \citep[similar to that in][]{2010ApJ...725..249S}, according to the $\Gamma$ distribution, the inner and outer boundaries of the BLR, and the eccentricity distribution of BLR clouds (assuming they are randomly distributed in the range from $0$ to $0.5$)\footnote{Our main results are not affected by assuming that all BLR clouds are on circular orbits or somewhat different eccentricity distributions.}, from which orbits of test particles can be randomly assigned. After transforming the initial conditions to the natural rotating coordinate system, we use the Dopri5 integration code \citep{Hairer93} to integrate the motion equations with an accuracy of $10^{-8}$. We generate $2$ million test particles and integrate every single particle for $500$ orbital periods, and find that the stable particles are about $98\%$.\footnote{We have checked that almost all particles left are stable by integrating $5000$ of them for further $10,000$ orbits.} We record the locations and velocities of the remaining stable particles of the last five orbital periods. Based on the above description, once the model parameters, i.e., mass ratio, bolometric luminosity, and semimajor axis of the BBH, are given, the orbits of discrete clouds in the BLR can be randomly generated according to the above Equations~\eqref{eq:Rga}-\eqref{eq:theta} and three-body simulations. Therefore, we can obtain the profile and the variation of its emission lines from these clouds. 

\subsection{Enhanced/Weakened Ionizing Flux Received by a Cloud}

The observed BEL is the superposition of line photons emitted from different clouds ionized by the central source. In the single BH case, the central source is fixed at the center and does not move, because the velocity of each cloud is relatively small and the ionizing photon flux it received from the central source should not be significantly Doppler boosted. For the case of BBH systems, however, it is not only the relative distance between a cloud and the central source that changes with time, but also the ionizing flux received by the cloud may also be significantly Doppler boosted due to the relativistic motion of the secondary BH. In these BBH cases, for a given cloud in the BLR, the projection of its relative velocity along the direction of its relative position to the central source is 
\begin{equation}
\beta_{||} =\frac{( \vec{\boldsymbol{v}}_{{\rm C}_i}- \vec{\boldsymbol{v}}_{\rm sec})\cdot( \vec{\boldsymbol{r}}_{{\rm C}_i}- 
\vec{\boldsymbol{r}}_{\rm sec} ) }  {|\vec{\boldsymbol{r}}_{{\rm C}_i}-\vec{\boldsymbol{r}}_{\rm sec}|}.  
\end{equation}
where $\vec{\boldsymbol{v}}_{{\rm C}_i}$ and $\vec{\boldsymbol{r}}_{{\rm C}_i}$ are the velocity and distance vectors of the BLR cloud to the mass center of the BBH system, and $\vec{\boldsymbol{v}}_{\rm sec}$ and $\vec{\boldsymbol{r}}_{\rm sec}$ are the velocity and distance vectors of the secondary BH to the mass center of the BBH system.
The DB factor for the ionizing flux received by this cloud is then
\begin{equation}
D_{{\rm C}_i2}= \left[\gamma_{{\rm C}_i2}(1-\boldsymbol \beta_{{\rm C}_i2,||})\right]^{-1}, 
\end{equation}
where $\gamma_{{\rm C}_i}=1/\sqrt{1-|\vec{v}_{{\rm C}_i} - \vec{v}_{\rm sec}|^2/c^2}$ is the Lorentz factor.
Similar to that for the observed continuum, the ionizing flux received by a cloud C$_i$ is then given by
\begin{equation}
F_{\nu,\rm C} \propto F_{\nu,\rm e}(\nu; t_{\rm in}^{'}) D_{{\rm C}_i2}^{3- \alpha} \frac{|\vec{r}_{{\rm C}_i}|^2}{|\vec{r}_{{\rm C}_i} -\vec{r}_{\rm sec}(t'_{\rm in}) |^2},
\label{eq:BBH_flux}
\end{equation}
where the term $D_{{\rm C}_i2}^{3-\alpha}$ represents the enhancement/weakening of the ionizing flux received by the cloud due to the relativistic motion of the secondary BH with respect to the cloud, and the term $|\vec{r}_{{\rm C}_i}|^2/|\vec{r}_{{\rm C}_i} -\vec{r}_{\rm sec}(t'_{\rm in}) |^2$ considers the variation of the ionizing flux due to the  position change of the source (secondary BH). 
If the intrinsic variation is negligible, i.e., $F_{\nu,\rm e}(\nu_{\rm, e}; t_{\rm in}^{'}) \sim \textrm{const}$, then it is in the BBH-DB scenario; if the intrinsic variation is much more significant compared with the DB effect, then it is in the BBH-IntDB scenario. 

If we set $\vec{v}_{\rm sec}=0$ and $a_{\rm BBH}=0$, then it reduces to the single BH case, for which the variation due to the DB effect is indeed negligible as $|\vec{v}_{\rm C}| \ll |\vec{v}_{\rm sec} |$ and $|\vec{r}_{{\rm C}_i}|^2/|\vec{r}_{{\rm C}_i} -\vec{r}_{\rm sec}(t'_{\rm in}) |^2=1$; therefore, the observed continuum variation can only be due to the intrinsic variation. 
 
\subsection{Line Profiles and Their Variations}
\label{subsec:lineprofile}

Once we set the accreting BBH/BH system, the continuum variation, and the structure, and the cloud distribution of the BLR, we can obtain the BEL profiles and their variations $L(v,t_{\rm obs})$ directly by the summation of the line photons emitted from all BLR clouds. The line profile is also affected by the relative position and velocity change of the central ionizing source, in addition to the intrinsic variation of the ionizing source and the BLR structure and cloud distribution, i.e.,
\begin{eqnarray}
L(v,t_{\rm obs})
& \propto & \left. \sum_{i=1}^{N_{\rm tot}} F_{\nu,\rm e}(\nu; t_{\rm in}^{'}) D^{3-\alpha}_{{\rm C}_i2}\frac{|\vec{r}_{{\rm C}_i}|^{2}}{|\vec{r}_{{\rm C}_i} - \vec{r}_{\rm sec}|^2}  \right|_{v= { v^{\rm tot}_{{\rm C}_i}}   }. 
\label{eq:BBHLp}
\end{eqnarray}
Here, $N_{\rm tot}$ represents the total number of BLR clouds, $v^{\rm tot}_{{\rm C}_i} \simeq v^{\rm D}_{{\rm C}_i} + v^{\rm g}_{{\rm C}_i} $ is the summation of the Doppler redshift/blueshift  $v^{\rm D}_{{\rm C}_i}=\left\{\left[ \gamma_{{\rm C}_i{\rm obs}}(1-\vec{\beta}_{{\rm C}_i{\rm obs}}\cdot \boldsymbol{\hat{l}})\right]^{-1}-1\right\} c$  and the gravitational redshift of line photons from individual BLR cloud received by a distant observer $v^{\rm g}_{{\rm C}_i} =  GM/(r_{\rm C_{i}}c) $   \citep[see observational evidence in][]{2014ApJ...794...49T}, $\vec{\beta}_{{\rm C}_i{\rm obs}}$ represents the relative velocity of the cloud C$_i$ relative to the distant observer in units of the speed of light, $\gamma_{{\rm C}_i{\rm obs}}=1/\sqrt{1-|\vec{\beta}_{{\rm C}_i{\rm obs}}|^2}$, 
 $M$ is the total mass of the central BH/BBH system, and $r_{\rm C_{i}}$ is the distance of the BLR cloud to the mass center of the BH/BBH system.
 
For a single BH source with intrinsic variation, then the above equation is reduced to 
\begin{eqnarray}
L(v,t_{\rm obs})
& \propto & \left. \sum_{i=1}^{N_{\rm tot}} F_{\nu,\rm e}(\nu; t'_{\rm in}-\tau_{{\rm C}_i})\right|_{ { v=v^{\rm tot}_{{\rm C}_i}}} \nonumber \\
& \propto& \int \psi(v,\tau) F_{\nu,\rm e}(\nu; t'_{\rm in}-\tau) d\tau,
\label{eq:BHLp}
\end{eqnarray}
where $\psi(v,\tau)$ is the traditionally defined two-dimensional transfer function (2DTF), which explicitly reflects the BLR structure and the distribution of clouds in the BLR.
One may note here, for the single BH case, that the line profile and its variation is directly a convolution of the continuum variation with the 2DTF, and the 2DTF is solely determined by the BLR structure and cloud distribution (Eq.~\eqref{eq:BHLp}). However, for the BBH cases, it is difficult to define such a continuum-independent 2DTF, because the position change of the secondary BH and the DB effects on the ionizing flux received by clouds lead to the entanglement between the structure and cloud distribution of the BLR and the continuum variation (Eq.~\eqref{eq:BBHLp}). These two effects will be analyzed separately in Section~\ref{subsec:analyse}. 

Given the variation of the ionizing continuum emission (including its dependence on the direction) and the BLR structure and kinematics, the profiles of emission lines and their variations can be obtained by using Equations~\eqref{eq:BBHLp} and \eqref{eq:BHLp}. 

\section{Model Settings}
\label{sec:modelset}

\begin{table*}
\centering
\caption{
Model Parameters for a BBH/BH System with a (Circumbinary) BLR under Different Scenarios for Optical/UV Periodicity.
}
\begin{tabular}{lcccccccccc} \hline 
\multirow{2}{*}{Model} & $M_{\rm tot}$ & Mass &\multirow{2}{*}{$T_{\rm orb}$} & $a_{\rm BBH}$ & \multirow{2}{*}{$\lambda_{\rm Edd}$} &  \multirow{2}{*}{$\mathcal{A}$} & \multirow{2}{*}{$\mathcal{A}_{\rm Int}$} & \multicolumn{3}{c}{BLR}\\ \cline{2-2} \cline{5-5} \cline{9-11}
                        & $(10^9M_{\odot})$ & Ratio & &($10^{-3}$\,pc) & & & & $R_{\rm BLR}$\,(pc)
                        & $i (^\circ)$ & $\theta_{\rm o} (^\circ)$ \\ \hline \hline
BBH-DB-li        & $5 $  & 0.2     &2       & 13     &0.09  & 0.28 & 0    & 0.06 & 30 & 30\\
BBH-DB$^{*}$-li  & $5 $  & 0.2     &2       & 13     & -    &0.28  & 0    & 0.6  & 30 & 30\\
BBH-IntDB-li     & $0.5$ & 0.2     &2       & 6    & 0.9    &0.28  & 0.13 & 0.06 & 30 & 30\\
BH-Int-li        & $0.5$ &$\cdots$ &$\cdots$ &$\cdots$ &0.15 & 0.28 &0.28 &0.06  & 30 & 30\\  \hline \hline

BBH-DB-hi       & $5$   & 0.2     &2       & 13     &0.09  &0.57  & 0    & 0.06 & 85 & 30\\
BBH-DB$^{*}$-hi & $5$   & 0.2     &2       & 13     & -    &0.57  & 0    & 0.6  &85 & 30 \\
BBH-IntDB-hi    & $0.5$ & 0.8     &2       & 6      &0.34   &0.57 &0.34  & 0.06  &85 & 30 \\
BH-Int-hi       & $0.5$ &$\cdots$ &$\cdots$ &$\cdots$ & 0.15 &0.57 & 0.57 & 0.06 & 85 & 30  \\ \hline
\end{tabular}
\begin{flushleft}
\tablecomments{
Columns from left to right list the model name, the (total) mass of the BBH/single BH $M_{\rm tot}$, the variation period of the continuum or the orbital period of the BBH system $T_{\rm orb}$, the BBH semimajor axis $a_{\rm BBH}$, the Eddington ratio of the BBH/single BH accretion system ($\lambda_{\rm Edd}$), the total variation amplitude $\mathcal{A}$ and the amplitude of the intrinsic variation $\mathcal{A}_{\rm Int}$, the mean BLR size $R_{\rm BLR}$, the viewing angle $i$, and the opening angle of the BLR, respectively.  
}
\end{flushleft}
\label{tbl:t1}
\end{table*}

In this section, we describe our detailed model settings for different scenarios. We are aiming to show the differences in the variations of BELs resulting from the BBH-DB, BBH-IntDB, and BH-Int systems by setting two types of model parameters as listed in Table~\ref{tbl:t1}. For all the systems, we assume the observed continuum variation is periodic with a period of $2$\,yr, for simplicity, as those periodic quasars found in \citet{2015MNRAS.453.1562G} and \citet{2016MNRAS.463.2145C} typically have periods of around a year to several years, and the semimajor axes of those BBH systems can be consequently obtained once the total mass and mass ratio are fixed.

For the first type of model (first four rows in Table~\ref{tbl:t1}), we assume that all different systems are all viewed at an inclination angle close to face on ($i=30^\circ$), and the observed continua vary periodically with (roughly) the same amplitude $\mathcal{A}=0.28$. The BH-Int-li model (fourth row) assumes a single central BH with mass $0.5\times 10^9M_\odot$ and mean Eddington ratio of $0.15$, typical of quasars. The BLR size is roughly $0.06$\,pc according to Equation~\ref{eq:RBLR}, and the opening angle of the BLR is assumed to be $30^\circ$.

For the BBH-DB-li system (first row), the continuum variation is solely caused by the (relativistic) motion of the secondary BH; for the BBH-IntDB-li system (third row), the variation is a combination of the intrinsic variation and that due to the Doppler-boosting effect, the contribution fractions of which can be estimated according to Equations~\eqref{eq: DBamp} and \eqref{eq:Intamp} for a given BBH system. For the BBH-DB system to have the same variation amplitude as the BH-Int system, we set the total mass of the BBH $M_{\rm tot} = 5\times 10^9M_\odot$  (substantially larger than that for the BH-Int-li) and mass ratio of $q=0.2$ (BBH-DB-li). The optical/UV continuum radiation is from a disk associated with the secondary BH with the same mean luminosity as the BH-Int system. The BLR is also set to be the same as that for the BH-Int system. Because the BBH-DB-li system has a much larger total mass than that of the BH-Int-li system but the same BLR size, the widths of the BELs from BBH-DB-li are substantially larger than those from BH-Int-li, and thus can be easily differentiated from BH-Int-li. Considering of this, we also set a BBH-DB*-li system, as a toy model, with most parameters the same as the BBH-DB system, except that its mean BLR size is forced to be roughly 10 times larger than that set for the BH-Int system (though violating the $R_{\rm BLR}-L$ relationship), thus the resulting BELs have roughly the same width as those from BH-Int-li. 
For the BBH-IntDB system (BBH-IntDB-li), the total mass is set to be the same as the BH-Int system, the mass ratio and mean Eddington ratio are set to be $0.2$ and $0.9$, thus the mean luminosity and BLR size are the same as those of BH-Int-li. In this case, the contribution from the intrinsic variation must be $\mathcal{A}_{\rm Int} \sim 0.13$ according to Equation~\eqref{eq:Intamp}. 

For the second type of models (fifth to eighth rows in Table~\ref{tbl:t1}), we assume that all the systems are viewed at an inclination angle close to edge on, i.e., $85^\circ$. For the detailed model parameter settings, the parameters of BH-Int-hi, BBH-DB-hi, BBH-DB*-hi are the same as those of BH-Int-li, BBH-DB-li, and BBH-DB*-li except that the inclination angle is set to $85^\circ$. While for the BBH-IntDB-hi system, the mass for the BBH-IntDB-hi system is the same as that for BBH-IntDB-li, but with a higher mass ratio, which leads to a slower orbital velocity of the secondary BH. In this case, the contribution from intrinsic variation is dominant ($\mathcal{A}_{\rm Int}=0.34$ in the BBH-IntDB-hi scenario. The DB-induced apparent continuum variation is $\mathcal{A}=0.57$ for these two systems is due to the close to edge-on viewing.

\section{Results }
\label{sec:results}

The main goal of this paper is to explore whether it is possible to distinguish the BBH-DB, BBH-IntDB, and BH-Int scenarios with similar observed continuum (periodic) variations through the corresponding variations of the BELs. In this section, we first demonstrate the response of BEL (both flux and shape) to the position change of the source (mainly the disk associated with the secondary BH), the DB by the (relativistic) motion of the secondary BH, and the intrinsic variation of the source, separately. Then we show the variations of the line profiles resulting from different scenarios, i.e., BBH-DB, BBH-IntDB, and BH-Int, by convolution of all these effects. Finally, we also investigate the dependence of the BEL variation on several model parameters, including the opening angle of the BLR, viewing angle, and different compositions of corotating and counterrotating BLR clouds.

\subsection{Dependence on position variation of the secondary BH}
\label{subsec:analyse}

\begin{figure*}[htp]
\centering
\includegraphics[width=0.95 \textwidth]{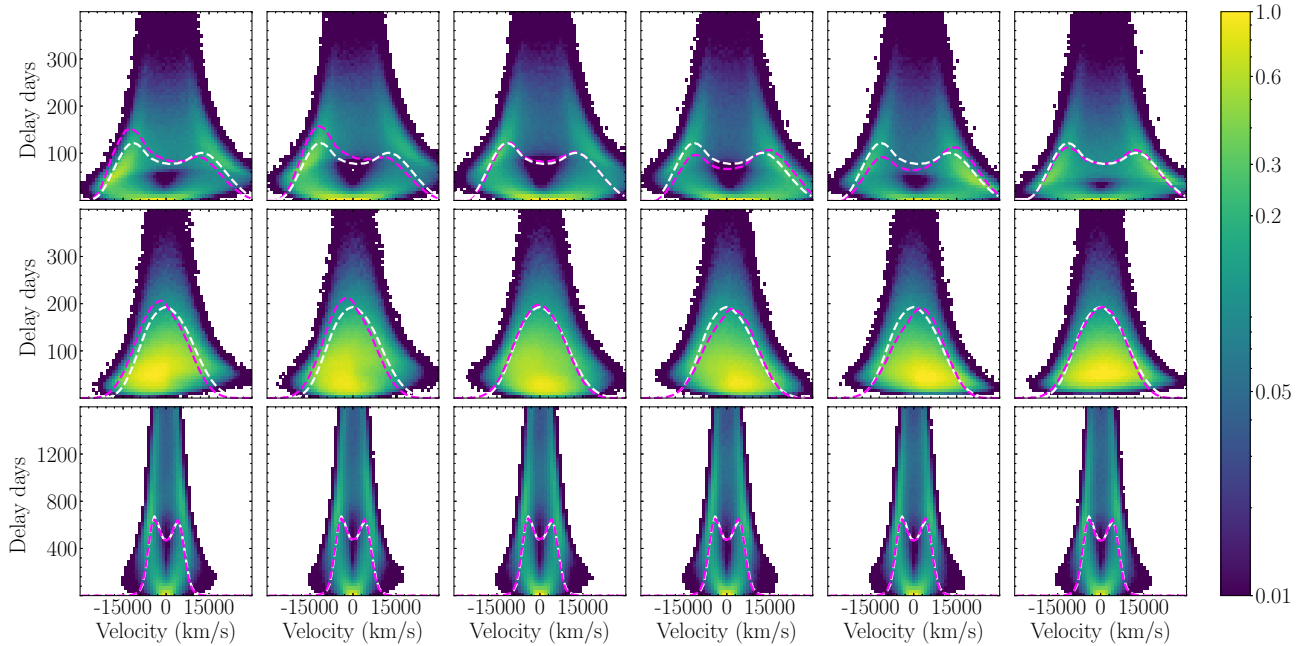}
\caption{Time delay versus velocity maps obtained by only considering the position variation of the secondary BH over a period. The top, middle, and bottom rows show the cases for a BBH-BLR system as listed in Table~\ref{tbl:t1}, i.e., the first (BBH-DB-hi), second (BBH-DB-li), and third (BBH-DB$^{*}$-hi) rows. Panels from left to right in each row show the results obtained at six different orbital phases of the secondary BH, i.e., $0$, $\pi/3$, $2 \pi/3$,  $\pi$, $4 \pi/3$, and $5\pi/3$, respectively. In each panel, the red dashed line shows the relative velocity profile at the moment of observation in arbitrary units, and the white line shows the mean profile over a whole orbital period. The color bar at the right side marks the relative brightness of each pixel compared to the brightest one in the first panel of each separate row. 
}
\label{fig:f2}
\end{figure*} 

The position variation of the secondary BH may affect the BEL shapes periodically if the $r_{{\rm C}}$ of many BLR clouds is not much larger than $a_{\rm BBH}$ (Eq.~\ref{eq:BBHLp}), say $\gg 10$. Figure~\ref{fig:f2} shows the time delay versus velocity maps by only considering the position variation of the secondary BH for a few BBH systems  
as examples to illustrate the effects of this position variation. 
The top, middle, and bottom rows of Figure~\ref{fig:f2} show the cases for a BBH-BLR system based on the parameters listed in Table~\ref{tbl:t1}, i.e., the fifth row (BBH-DB-hi), first row (BBH-DB-li), and sixth row (BBH-DB$^{*}$-hi) of Table~\ref{tbl:t1}.
These three examples have $R_{\rm BLR}/a_{\rm BBH}= 4.6$ (top and middle panels) and $46$ (bottom panel). In each panel, the magenta dashed line represents the velocity profile with arbitrary units (integration of the 2D map over time delay) by only considering the position change of the secondary BH as the source of the ionizing continuum, and the white dashed line represents the mean profile over a whole orbital period. As clearly seen from this figure, the map may change significantly with the position change of the secondary BH and so does the line profile, especially for the top row (close to edge on), as the term $\frac{|\vec{r}_{{\rm C}_i}|^2}{|\vec{r}_{{\rm C}_i} -\vec{r}_{\rm sec}(t'_{\rm in}) |^2}$ in Equation~\eqref{eq:BBH_flux} may vary widely from $0.67$ to $1.63$ ($R_{\rm BLR}/a_{\rm BBH}= 4.6 $); the variation is visible but less significant for the case with a small viewing angle (close to face on; middle panel), because the velocities for BLR clouds are significantly suppressed due to the face-on orientation and flattened BLR. When $R_{\rm BLR}/a_{\rm BBH}$ is much larger, e.g., $46$, the term $\frac{|\vec{r}_{{\rm C}_i}|^2}{|\vec{r}_{{\rm C}_i} -\vec{r}_{\rm sec}(t'_{\rm in}) |^2}$ varies much less significantly, e.g., roughly from $0.96$ to $1.04$, 
and the effects of the position change of the secondary BH can hardly be seen in the bottom row of Figure~\ref{fig:f2}. 

\subsection{Effects of Doppler Boosting and Intrinsic Variation on Ionizing Flux Received by BLR Clouds}

The enhancement/weakening of the ionizing flux received by BLR clouds at different positions in the BBH-DB scenario (without intrinsic variation) is quite different from that in the BBH-IntDB scenario. The main reason is that in the BBH-DB scenario, the enhancement occurs in the part of BLR roughly around the direction of motion of the secondary BH and the weakening occurs in the part of the BLR roughly around the opposite direction (but modified by the time-delay effect), while in the BBH-IntDB scenario, if only the intrinsic variation is considered but not the DB effect (hereafter BBH-Int(DB)), the enhancement and weakening occur at all directions though are also modified by the time-delay effect. 

\begin{figure*}
\centering
\includegraphics[width=0.95\textwidth]{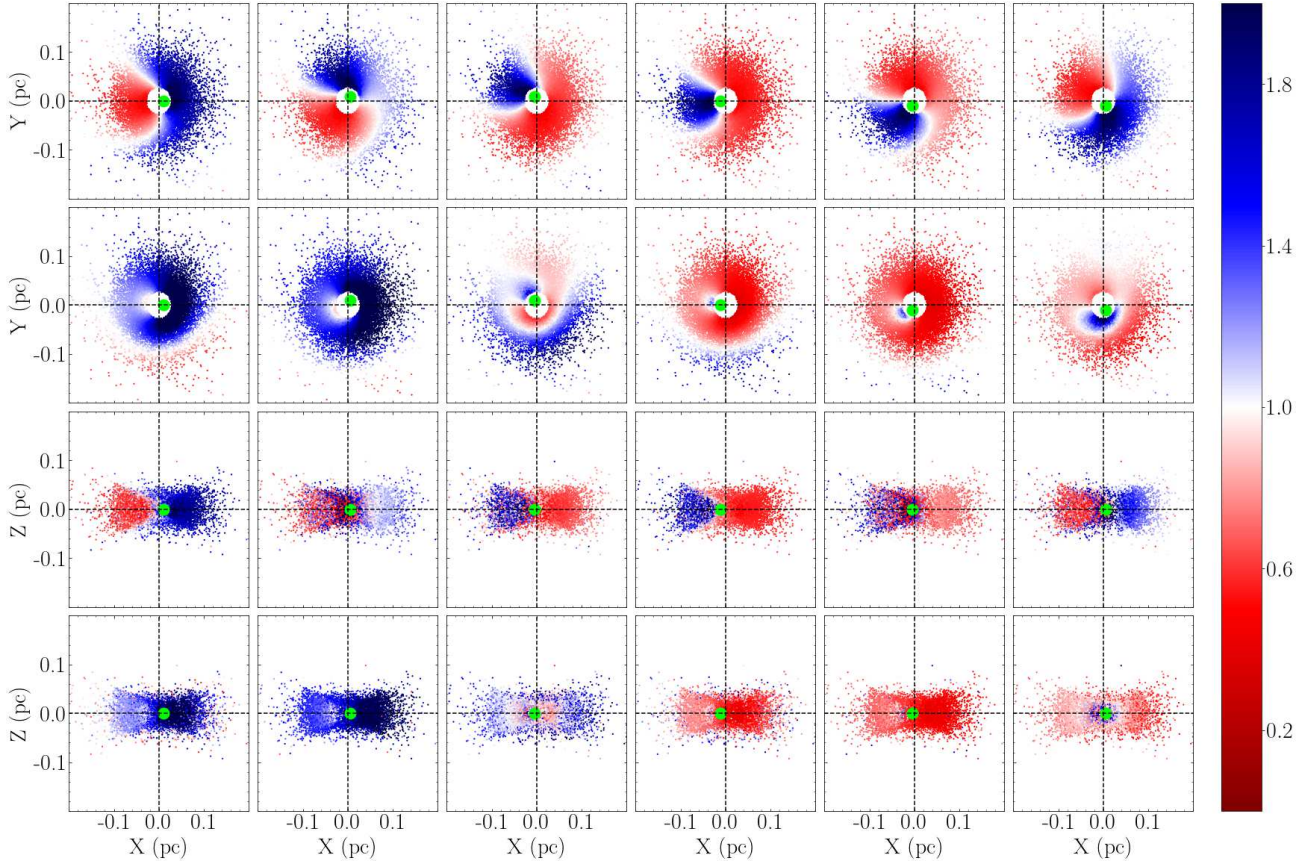}
\caption{
Enhancement/weakening of ionizing flux received by individual BLR clouds relative to the intrinsic ones, projected on the $xy$ plane (top two panels) and the $x-z$ plane (bottom two panels) at different phases of the periodic optical/UV continuum variations for the BBH-DB scenario (first and third rows) and the BBH-Int(DB) scenario (second and fourth rows). Note here that the parameters for the BBH-DB and BBH-Int(DB) systems shown here are based on the BBH-DB-hi model listed in Table~\ref{tbl:t1} (third/seven), except that the Doppler effect is ignored in the BBH-Int(DB) system for comparison.
Panels from left to right show the cases with continuum variation at phases of  $0$, $\pi/3$, $2\pi/3$, $\pi$, $4\pi/3$, and $5\pi/3$, respectively. As seen from top two panels, the ionizing- flux-enhanced and -weakened regions of the BLR in the BBH-Int(DB) scenario is almost symmetric (second row), which is different from the case with optical/UV variations induced by the DB effect (first row). The bottom two panels represent the time-delay effects in the BBH-DB (third row) and BBH-Int(DB) (fourth row) scenarios, respectively. The green points in the each panel represent the location of the secondary BH at different moments of observation. 
The color bar at the right side marks the relative flux received by each cloud (represented by each point) in the BLR in units of the mean flux at each given moment of observation. 
}
\label{fig:f3}
\end{figure*}

Figure~\ref{fig:f3} shows the spatial distribution of the enhancement/weakening of the ionizing flux received by individual clouds with respect to the intrinsic ionizing flux, projected onto the $xy$ plane (top two rows) and  $xz$ plane (bottom two rows). The basic parameters of the BBH system are similar to the BBH-DB-hi model in the seventh row of Table~\ref{tbl:t1}. In the first and third rows of Figure~\ref{fig:f3}, the intrinsic ionizing continuum is assumed to be constant and the observed periodic continuum variation is solely due to the DB effect, while in the second and bottom rows, the observed continuum variation is assumed to be solely due to the intrinsic variation and the DB effect is ignored (hereafter denoted as BBH-Int(DB)), for the purpose of showing clearly the difference between the DB effect and that due to intrinsic variation. As seen from the first/third and second/fourth rows of this figure, the enhancement/weakening pattern of the ionizing flux received by BLR clouds in the BBH-DB scenario is significantly different from that in the BBH-Int(DB) scenario. The former one is asymmetric and the enhancement/weakening region is rotating with the secondary BH and propagates outward due to the time-delay effect. The latter is closer to symmetric about the $y$-axis at the outer BLR region, though at the central BLR region, the enhancement/weakening part changes and rotates because of the position variation of the secondary BH. In such a case, the enhancement/weakening pattern also propagates outward periodically due to the time-delay effect. These different patterns for the enhancement/weakening of the ionizing flux received by BLR clouds will result in different behaviors of the periodic variation of BELs and their profile as detailed below.

\subsection{Variation of BELs and Their profiles}

In this subsection, we show the line profiles and their variations resulting from different systems with the same inclination angles. For the BBH-DB systems, we assume that the variation of
the observed continuum arises from the DB of a constant intrinsic continuum. For the BBH-IntDB system, the observed variation of the continuum is caused by the combination of both the DB effect and intrinsic variation of the source, and the intrinsic variation is significant or even dominant. For both BBH-DB and BBH-IntDB cases, the effect due to the position change of the secondary BH is also considered (see Section~\ref{subsec:analyse}) when generating the line profiles. While for the BH-Int system, the variation is only caused by the intrinsic variation of the central source. 

\begin{figure*}
\centering
\includegraphics[width=0.95\textwidth]{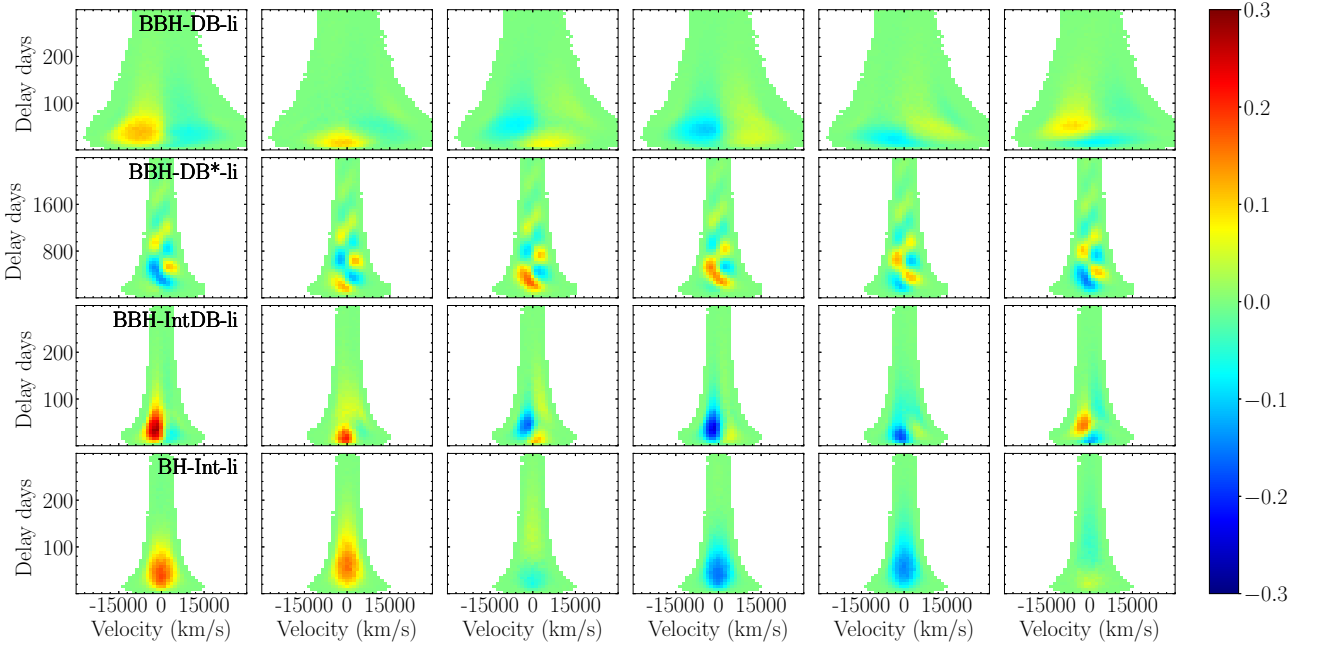}
\caption{
Differences of the time delay versus  velocity maps at different phases of the continuum variation from the mean map by considering the enhancement/weakening of the ionizing flux received by each cloud. Rows from top to bottom show the difference maps for the BBH-DB-li, BBH-DB*-li, BBH-IntDB-li, and BH-Int-li scenarios, the top four rows listed in Tab.~\ref{tbl:t1}, respectively. Panels from left to right shows the difference maps at phases $0$, $\pi/3$, $2\pi/3$, $\pi$, $4\pi/3$, and $5\pi/3$, respectively. The color bar at the right side indicates the relative flux variation in units of mean BEL flux at each separate row over a whole period of the continuum variation by multiplying an arbitrary scaling factor to set the color bar.
}
\label{fig:f4}
\end{figure*}

\begin{figure*}[htp]
\centering
\includegraphics[width=0.95\textwidth]{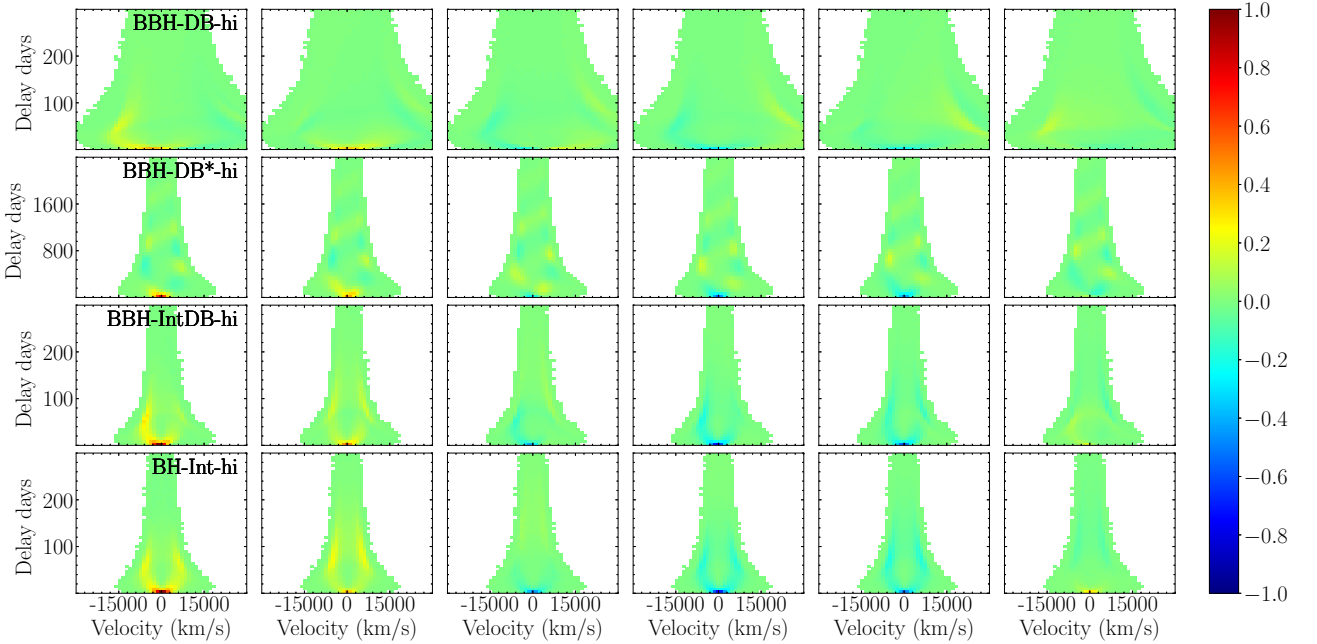}
\caption{ 
Legend similar to Fig.~\ref{fig:f4}, but rows from top to bottom represent the BBH-DB-hi, BBH-DB*-hi, BBH-IntDB-hi, and BH-Int-hi models, as the fifth to eighth rows listed in Tab.~\ref{tbl:t1}, respectively. 
}
\label{fig:f5}
\end{figure*}

The time delay versus velocity map defined in the present paper (different from the traditionally defined 2DTF), which depends on the changes of the position and direction of motion of the source (secondary BH) and thus phase dependent. To clearly show the variation of such maps with the periodic variation of the observed continuum, we plot in Figure~\ref{fig:f4} the differences between maps at different given phases and the mean map over the whole variation period. The top to bottom rows in Figure~\ref{fig:f4} show the results for the BBH-DB-li, BBH-DB*-li, BBH-IntDB-li, and BH-Int-li models, respectively. Panels from left to right show the map at the continuum variation phases of $0$, $\pi/3$, $2\pi/3$, $\pi$, $4\pi/3$, and $5\pi/3$, respectively. Similar to Figure~\ref{fig:f4}, Figure~\ref{fig:f5} shows the results for those systems listed in the fifth to eighth rows in Table~\ref{tbl:t1}, respectively. 

\begin{figure*}
\centering
\includegraphics[width=0.95\textwidth]{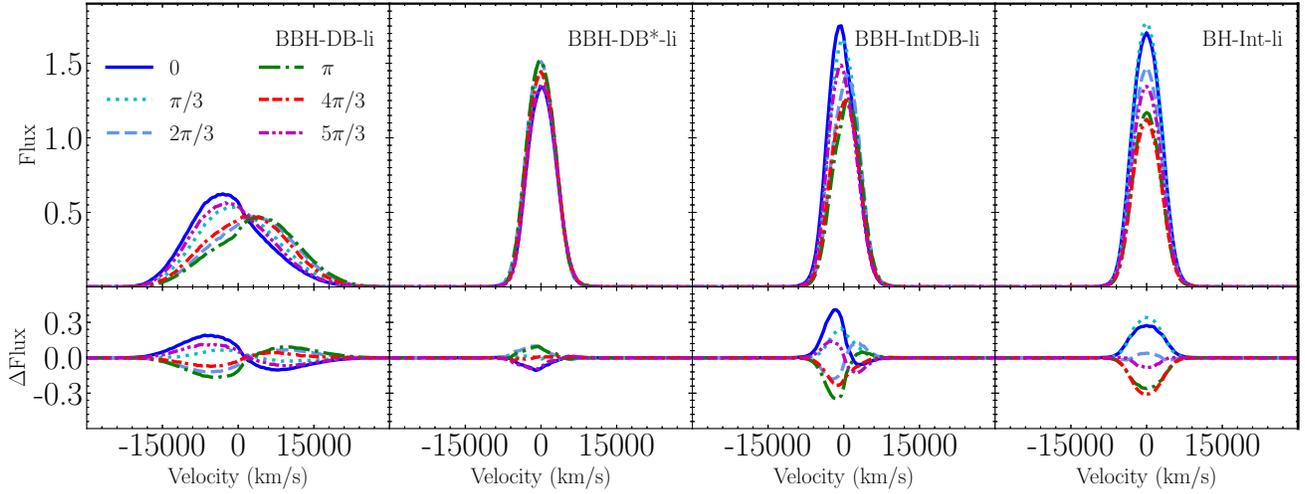}
\caption{
Variations of a BEL over a single period of continuum variation, resulting from the BBH-DB-li (left column), BBH-DB*-li (second column), BBH-IntDB-li (third column), and BH-Int-li (right column) systems, respectively, same as those shown in Fig.~\ref{fig:f4}. The line flux is in arbitrary units. The top panels show the line profiles at six different phases, i.e., $0$ (blue line), $\pi/3$ (cyan dotted line), $2\pi/3$ (light blue dashed line), $\pi$ (green dashed-dotted line), $4\pi/3$ (red dashed-dashed-dotted line), and $5\pi/3$ (magenta dashed-dotted-dotted line), respectively, while the bottom panels show the difference of the line profile at each phase from the mean line profile, respectively. The model parameters are the same as those in Fig.~\ref{fig:f4}.
}
\label{fig:f6}
\end{figure*}

For the BBH-DB systems, as seen from the top row in Figures~\ref{fig:f4} and \ref{fig:f5}, the difference map at a given time is significantly asymmetric and there are regions in the blue part (negative velocities) and red part (positive velocities), which are significantly enhanced or weakened, due to both the DB effect and the position change of the ionizing source (associated with the secondary BH). The enhanced/weakened regions shift periodically with the phase of the observed continuum variation (i.e., the orbital phase in this case; left to right panels). Such variation in the time delay versus velocity map leads to a periodic variation of the BEL profiles (see Fig.~\ref{fig:f6}).  For the BBH-DB* system (second row in Figs.~\ref{fig:f4} and Fig.~\ref{fig:f5}), the difference maps are quite different from those shown for the BBH-DB systems simply because the BLR size relative to the BBH size is much larger than those of BBH-DB systems. The propagation of the continuum peak throughout the BLR can be clearly seen, as indicated by the magenta-blue features in the figure. For the BH-Int systems (the fourth row in Figs.~\ref{fig:f4} and \ref{fig:f5}), the difference maps are close to symmetric, which reflects the assumed isotropic intrinsic continuum radiation, although the enhanced/weakened region also shifts and changes periodically due to the periodic intrinsic continuum variation. For the BBH-IntDB systems (third row in Figs.~\ref{fig:f4} and \ref{fig:f5}), the difference map still shows significant asymmetry because of both the DB effect and the position change of the ionizing source, although the isotropic intrinsic variation leads to the weakening of such asymmetry. 

\begin{figure*}
\centering
\includegraphics[width=0.95\textwidth]{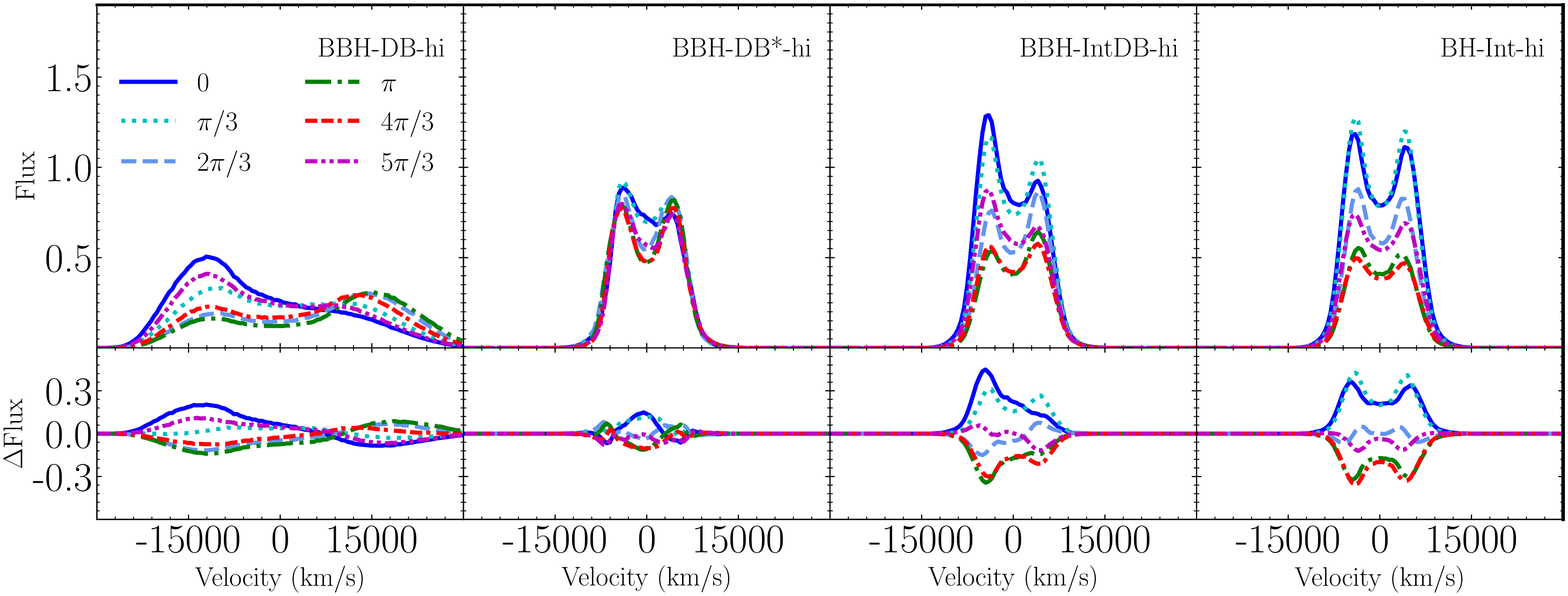}
\caption{Legend similar to Fig.~\ref{fig:f6}, but for the systems shown in Fig.~\ref{fig:f5}.
}
\label{fig:f7}
\end{figure*} 

Figures~\ref{fig:f6} and \ref{fig:f7} show the variations of the BEL profiles over a single period of the continuum periodic variation, resulting from those systems shown in Figures~\ref{fig:f4} and \ref{fig:f5}, respectively. In each panel of the top row, lines with different colors represent the BEL profile at different phases: $0$ (blue line; at which the distant observer receives the maximum continuum flux), $\pi/3$ (cyan dotted line), $2\pi/3$ (light blue dash line), $\pi$ (green dashed-dotted line), $4\pi/3$ (red dashed-dashed-dotted line), and $5\pi/3$ (magenta dashed-doted-dotted line), respectively. In each panel of the bottom row, lines with different colors show the difference between the BEL profile at each given phase and the mean BEL profile over a period of the continuum variation. Note that the mean BEL flux over a whole period of the continuum variation is normalized to $1$, and the BELs at different phases are all scaled by such a normalization.

For the BBH-DB systems (left columns in Figs.~\ref{fig:f6} and \ref{fig:f7}), the relative variations of the red and blue parts of the line are in opposite directions, which is expected as the parts of the BLR being lightened by the enhanced/weakened ionizing flux rotate periodically due to the orbital motion of the secondary BH and the DB effect. The change at the blue side of the BEL is more significant than the red part, which is mainly due to the asymmetric nature of the DB effect and the effect caused by the position change of the secondary BH. 
For the BBH-DB* systems (second columns in Figs.~\ref{fig:f6} and \ref{fig:f7}), the variation of the BELs becomes much weaker as the effect due to the continuum variation of BELs is averaged out due to the larger BLR size compared with the light traveling scale in an orbital period in this case. However, the DB effect can still lead to significant asymmetry in the variation of BELs at blue and red sides. The BEL variation can be even more complicated for the case with $i=85^\circ$ because of the much larger BLR size  and the wider velocity separations between enhanced  and weakened BLR regions compared with those in the BBH-DB models (see also Figs.\ref{fig:f4} and \ref{fig:f7}). 

For the BBH-IntDB systems, the BEL variations are caused by the combination of the DB effect and the intrinsic and position variations of the ionizing source. The asymmetry of the BEL variation at the red and blue sides becomes weaker, but the pattern of variation becomes more complicated due to the additional effect from the intrinsic variation (leading to symmetric BEL changes). We also note here that the much wider line widths of the BBH-DB models compared with that from the BBH-IntDB and BH-Int models are due to the masses of the BBH-DB systems are ten times larger than those of the BBH-IntDB or BH-Int systems, but the BLR sizes, roughly determined by the luminosity, are the same. 

For the BH-Int systems shown in the right columns in Figures~\ref{fig:f6} and \ref{fig:f7}, the BEL variation is almost symmetric about the zero velocity, and the red and blue wings increase/decrease simultaneously. The reason is that the ionizing flux with intrinsic variation is assumed to be isotropic, and the central source is at rest. This variation pattern is clearly different from the BBH cases shown in the same figures, which can be taken as the signature to distinguish the BBH system from the single BH system.
It is also clear as described above (Figs.~\ref{fig:f6} and \ref{fig:f7}) that the variation pattern of the BEL profiles resulting from systems with variations dominated by the DB effect is different from that from systems with variation dominated by the intrinsic ones. Based on such a difference, one should be able to distinguish BBH-DB systems from BBH-IntDB systems.

\subsection{Dependence on the BLR inclination angles }

In the previous subsection, we compare the BEL variations for two sets of parameters, as listed in Table~\ref{tbl:t1}, with the inclination angles to be $i=30^{\circ}$ and $85^{\circ}$, respectively. In real observations of such BBH candidates with periodic optical/UV variations, the inclination angle $i$ is not known at first. Therefore, it would be interesting to investigate the dependence of BEL variations on $i$ for a given BBH system under the condition that the observed continuum variation is the same. In this case, the contribution fractions from the DB effect and the intrinsic variations change with $i$, as already mentioned above. When $i$ is small, the DB effect would be less significant and the required intrinsic variation would be stronger, while the DB effect would be much more significant when $i$ is close to $90^\circ$. Note that this is not the same as the situation where the same BBH system with fixed (or zero) intrinsic continuum variation is viewed from different inclination angles.

\begin{figure*}[htp]
\centering
\includegraphics[width=0.95\textwidth]{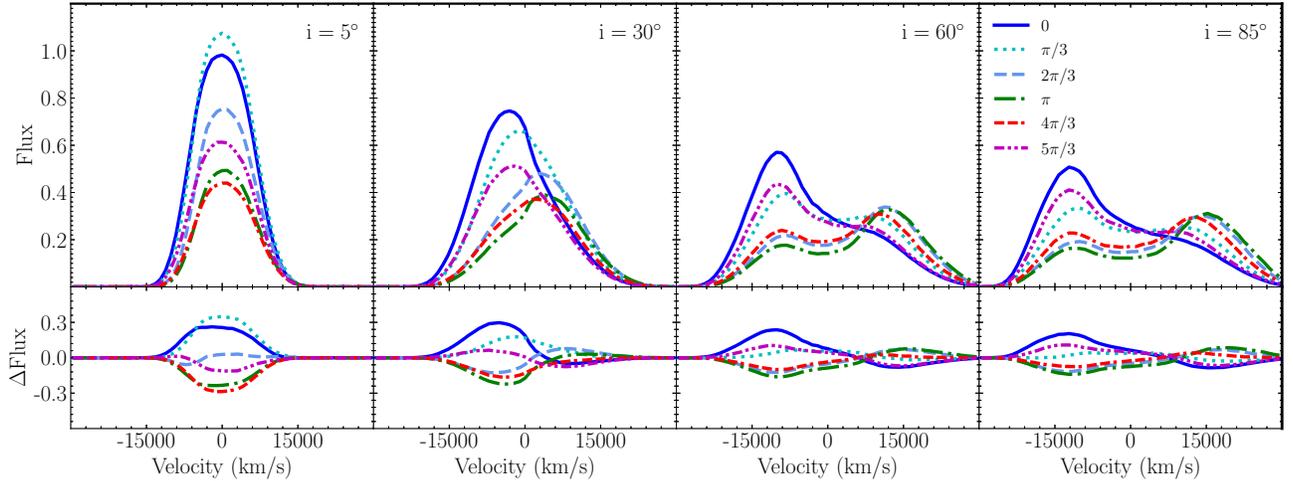}
\caption{ Variations of profiles over a period of the continuum variation for a BBH system with the BLR viewed at different inclination angles. The left to right columns show the cases with inclination angles of $5^{\circ}$, $30^{\circ}$, $60^{\circ}$, and  $85^{\circ}$, respectively. Each panel in the top row shows the line profile shows the line profiles at six different phases, i.e., $0$ (blue line), $\pi/3$ (cyan dotted line), $2\pi/3$ (light blue dashed line), $\pi$ (green dashed-dotted line), $4\pi/3$ (red dashed-dashed-dotted line), and $5\pi/3$ (magenta dashed-dotted-dotted line), respectively, while each panel in the bottom row shows the difference of the line profile at each phase from the mean line profile, respectively. The parameters of the BBH system are based on the BBH-DB-hi in the bottom panel of Tab.\ref{tbl:t1} under the condition of different inclination angles.
}
\label{fig:f8}
\end{figure*} 

Figure~\ref{fig:f8} shows line profile variations for the BBH system with parameters the same as BBH-DB-hi in Table~\ref{tbl:t1}, except that $i$ is set as $5^{\circ}$, $30^{\circ}$, $60^{\circ}$, and $85^{\circ}$  from left to right columns, respectively. As seen from this figure that the double-peaked BEL profiles (right two columns for close to edge-on BLR orientation, $i=85^\circ$ and $60^\circ$, with $(\mathcal{A_{\rm Int}}, \mathcal{A_{\rm DB}})= (0, 0.57)$ and $(0.053, 0.49)$) change periodically, with the red and blue wings varying in the opposite directions to the DB effect, and the effect due to the position change of the central source dominates the ionizing flux variation. If the inclination angle is small (nearly face on, e.g., the left column with $i=5^\circ$), the continuum variation is dominated by the intrinsic ones ($(\mathcal{A_{\rm Int}}, \mathcal{A_{\rm DB}})=(0.51, 0.037)$), and consequently, the variation of the Gaussian-like BEL profiles becomes close to symmetric. The case for $i=30^\circ$ (second column; $(\mathcal{A_{\rm Int}}, \mathcal{A_{\rm DB}})=(0.23, 0.28)$) is in between the close to face-on ones (left column) and the close to edge-on ones (right two columns).

\subsection{Dependence on BLR opening angles}

The BEL profiles depend on the opening angle of the BLR, and so its variation may also depend on it. In the above figures, we only show those cases with fixed a BLR opening angle ($\theta_{\rm o}=30^\circ$). Below we illustrate the dependence of the BEL variations on the BLR opening angle.

\begin{figure*}
\centering
\includegraphics[width=0.95\textwidth]{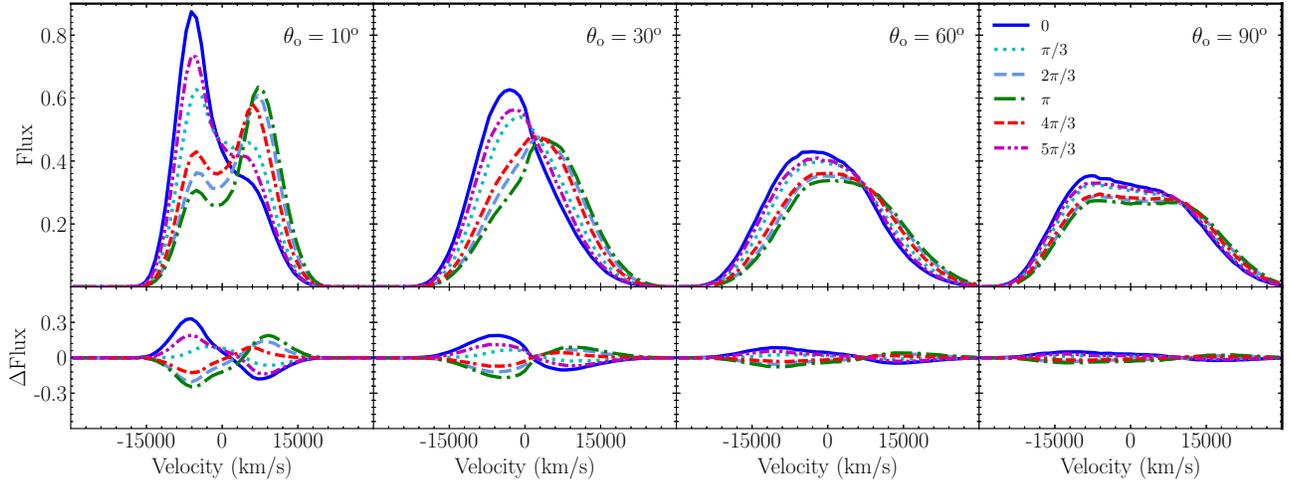}
\caption{
Variations of BEL profiles over a single period for the BBH-DB scenario at different opening angles. Columns from left to right show the results for BLR opening angles $\theta_{\rm o} = 10^\circ$, $30^\circ$, $60^\circ$, and $90^\circ$ (spherical). Each panel in the top row show the line profiles at phases $0$ (blue line), $\pi/3$ (cyan dotted line), $2\pi/3$ (light blue dash line), $\pi$ (green dashed-dotted line), $4\pi/3$ (red dashed-dashed-dotted line), and $5\pi/3$ (magenta dashed-dotted-dotted line), while the corresponding bottom panel shows the difference of the line profile at each phase from the mean line profile, respectively. The BBH-DB system is constructed based on the BBH-DB-li model as listed in Table~\ref{tbl:t1}, but with the opening angle set to $10^{\circ}, 30^{\circ}, 60^{\circ}$, and $90^{\circ}$, respectively.
}
\label{fig:f9}
\end{figure*}

\begin{figure*}
\centering
\includegraphics[width=0.95\textwidth]{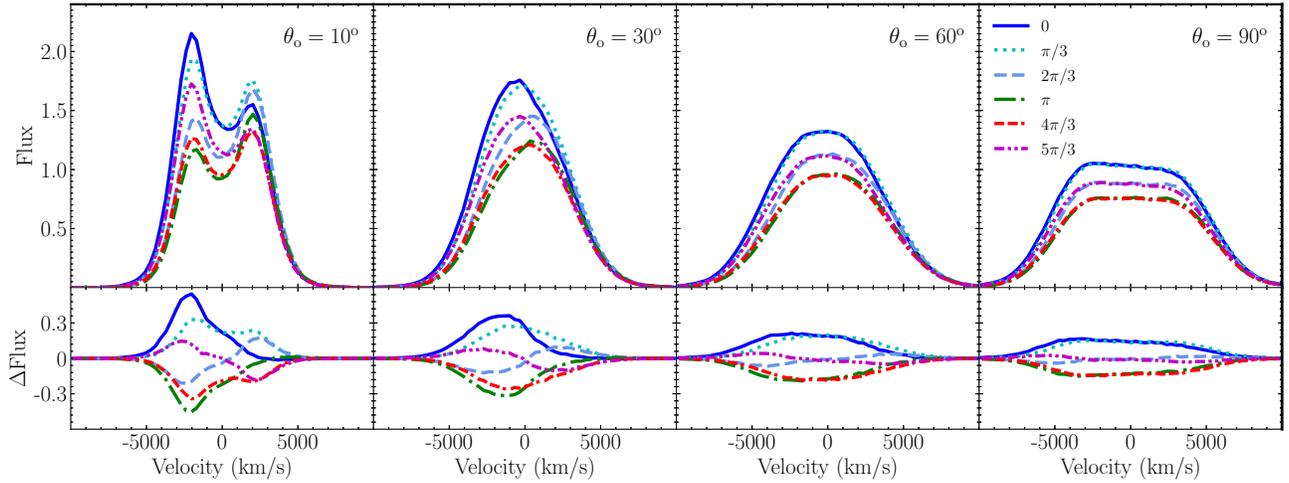}  
\caption{Variations of BEL profiles over a single period for the BBH-IntDB scenario at different opening angles, with legend similar to Fig.~\ref{fig:f9}. The BBH-IntDB system is constructed based on the BBH-IntDB-li model, but with the opening angle set to $10^{\circ}, 30^{\circ}, 60^{\circ}$, and $90^{\circ}$, respectively.
}
\label{fig:f10}
\end{figure*}

Figures~\ref{fig:f9} and \ref{fig:f10} show the BEL profile variations for BBH-DB and BBH-IntDB systems with parameters the same as those of BBH-DB-li and BBH-IntDB-li models listed in Table~\ref{tbl:t1}, except that the BLR opening angle is set to $\theta_{\rm o} = 10^\circ$, $30^\circ$, $60^\circ$, and $90^\circ$, respectively. For a BLR with a flattened disk-like structure ($\theta_{\rm o} = 10^\circ$), the line profiles at different phases present double-peaked features even though they are viewed at an inclination angle of $i=30^\circ$. With increasing $\theta_{\rm o}$, the BEL profiles become Gaussian-like at $\theta_{\rm o} \sim  30^\circ-60^\circ$ and finally top flat for spherical BLRs. The phase-dependent profile variations for the BBH-DB and BBH-IntDB scenarios are clearly seen from the bottom rows of Figures~\ref{fig:f9} and \ref{fig:f10} for the relative flux variation ($\Delta \rm Flux$) from the mean BEL profile. 

For the BBH-DB model, the BEL profile variation is caused by the DB effect and the position change of the secondary BH (the ionization source), while for the BBH-IntDB model, the intrinsic variation adds an additional effect to the BEL profile variation. We summarize the difference in the BEL profile variations between these two models, mainly the following two aspects.
\begin{itemize}
\item For the BBH-DB model, the maximum fluctuation amplitude appears in the most flattened disk-like BLR, e.g., $\theta_{\rm o} = 10^\circ$, as seen from Figure \ref{fig:f9}. Because those BLR clouds with an orbital plane perpendicular to the BBH orbital plane are less affected than those parallel to it by the DB effect, the relative flux variations of  BELs resulting from spheroid-like BLRs are smaller than those from flattened ones. For the BBH-IntDB model, the response of the line emission from BLR clouds at high inclination angles to the intrinsic ionizing continuum variation is the same as that of the clouds at low inclination angles, and because the intrinsic variation dominates the total continuum variability, therefore, the BEL amplitude variation is less significantly affected by the choice of the BLR opening angle (see Fig.~\ref{fig:f10}).
\item For the BBH-DB model, $\Delta {\rm Flux}$ goes to the maximum and minimum peaks at the blue and red velocity-shifted centers alternatively, with $\Delta {\rm Flux} \sim 0$ at a small positive velocity (close to zero when $\theta_{\rm o}=10^\circ$ and $30^\circ$), independent of the BEL profiles, whether double-peaked, Gaussian-like, or top-flat ones (see the bottom row in Fig.~\ref{fig:f9}). For the BBH-IntDB model, the variation of $\Delta{\rm Flux}$ is different from the case in the BBH-DB model (bottom row in Fig.~\ref{fig:f10}); when the blue side is at the maximum, the red side can also be close to maximum. This pattern difference of the BEL line profile changes resulting from these two models indicates the generic difference of the continuum variations. In the former one, the continuum variation is  apparent and it is due to the DB effect of the periodic motion of the secondary BH, thus the blue and red parts of the BELs are affected periodically one after the other. However, in the latter one, the blue and red parts of BELs are affected by the intrinsic continuum variation at the same time. If the intrinsic flux variation dominates, and the DB effect is negligible,  the blue and red parts of BELs would be simultaneously enhanced or weakened and change symmetrically.
\end{itemize}

\subsection{ Dependence on Co- and Counterrotating BLR Clouds}

\begin{figure*}[htbp]
\centering
\includegraphics[width=0.95\textwidth]{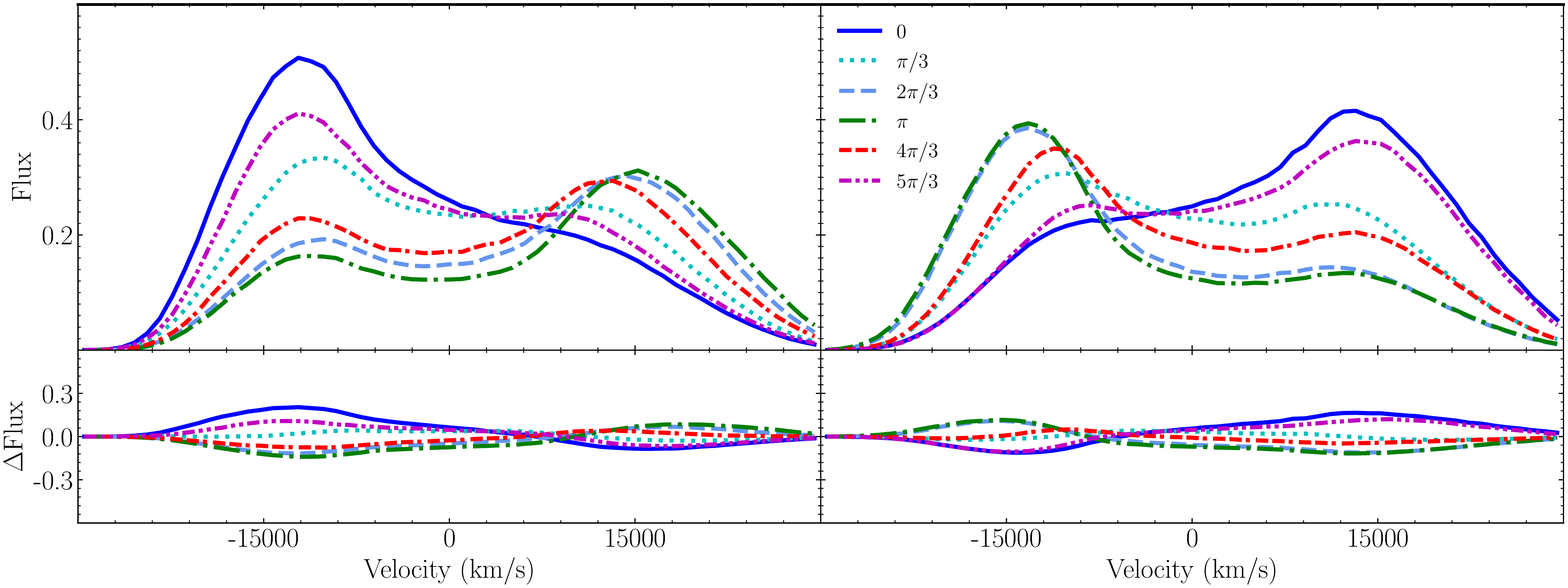}
\caption{
Variations of the BEL profiles resulting from the BBH-DB models with BBHs viewed edge on. The left and right panels show the results for the BBH-DB systems with all BLR clouds corotating and counterrotating, respectively. In each top panel, the variations are presented by profiles at six phases (0, $\pi/3$, $2\pi/3$, $\pi$, $4\pi/3$, $5\pi/3$) over a single continuum variation/orbital period, with the difference of the profile at each phase from the mean profile shown in the corresponding bottom panel. Parameters for the BBH-DB systems shown in this figure are the same as those for the BBH-DB-hi model listed in Table~\ref{tbl:t1}, except that the BLR clouds are all set to be corotating in the left panels, the same as those for the BBH-DB-hi model, but counterrotating in the right panels.
}
\label{fig:f11}
\end{figure*} 

\begin{figure*}[htbp]
\centering
\includegraphics[width=0.95\textwidth]{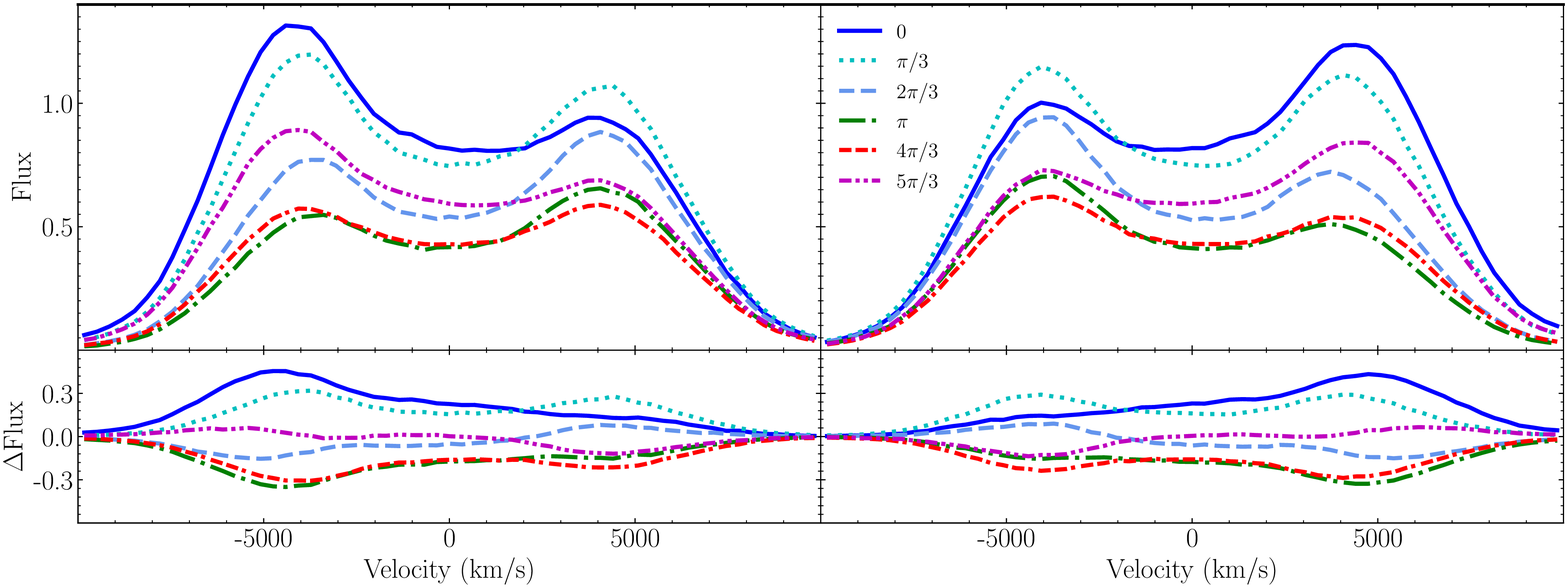}
\caption{
Legend similar to Figure~\ref{fig:f11}, but for the BBH-IntDB models with parameters the same as those for the BBH-IntDB model, except that the BLR clouds are all set to be corotating with the central BBH system in the left panels, the same as that for the BBH-IntDB-hi model, but counterrotating in the right panels.
}
\label{fig:f12}
\end{figure*} 

In the above calculations and analyses, we assume that all the clouds in the BLR are corotating with the BBH. Note that the assumption of the same rotation direction for all BLR clouds is compatible with the observations of 3C 273 by GRAVITY on board  the Very Large Telescope Interferometer \citep[VLTI;][]{2018Natur.563..657G}, with which the BLR of 3C 273 and its rotation were resolved. However, the possibility of counterrotating clouds in the general case is not ruled out, especially in the case of a circumbinary BLR formed by the merger of two progenitor massive black hole systems. Here we further investigate how the variations of BELs depend on the rotation direction of BLR clouds. We set two different cases with either all BLR clouds corotating or counterrotating around the central BBH for both the BBH-DB and BBH-IntDB models, and the resulting BEL variations are shown in the left and right panels of Figures~\ref{fig:f11} and \ref{fig:f12}, respectively.  The parameters for the BBH-DB and BBH-IntDB models adopted here are the same as those for the BBH-DB-hi and BBH-IntDB-hi models, respectively.

\begin{figure*}
\centering
\includegraphics[width=0.95\textwidth]{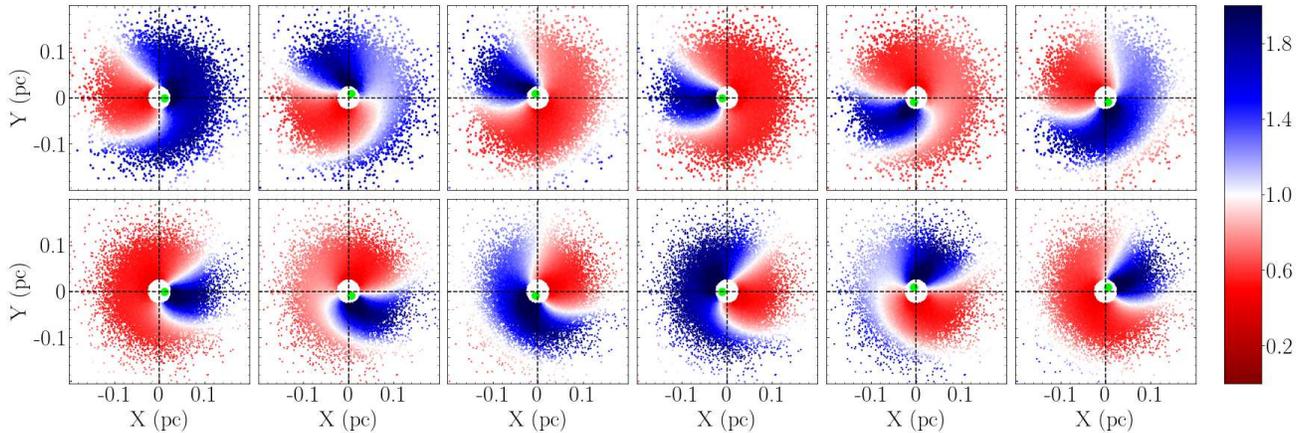}
\caption{
Legend similar to that for the top two rows of Figure~\ref{fig:f3}. For comparison, the top row of this figure is the same as the top row in Figure~\ref{fig:f3} (left panel of Fig.~\ref{fig:f11}), while the bottom row is for a BBH-DB system with the same parameters as those shown in the top row, but with BLR clouds counterrotating with the BBH (right panel of Fig.~\ref{fig:f11}).
}
\label{fig:f13}
\end{figure*}

The BEL profiles resulting from the BBH-DB model (Fig.~\ref{fig:f11}) with counterrotating BLR clouds (right panels) are different from those with corotating BLR clouds, and such difference may be used to obtain observational constraints on the rotating direction of BLR clouds. For a corotating-only BLR configuration, the BEL blue/red peak is close to the maximum/minimum value when the observed continuum is around the peak luminosity (phase $0$), and it changes to the minimum/maximum value when the observed continuum is around the lowest luminosity (phase $\sim \pi$; see the left column). If the BLR clouds are all counterrotating, then the BEL blue/red peak is close to the minimum/maximum value when the observed continuum is around the maximum luminosity (phase $\sim 0$), and it changes to the maximum/minimum value when the observed continuum is around the lowest luminosity (phase $\pi$; see the right column). Note also now that the variation amplitude of the red peak becomes larger than that of the blue peak, in contrast to the case where all BLR clouds are corotating. This is due to the combination of the DB effect with the time-delay effect as the DB affects the BLR at the opposite directions for the corotating and counterrotating BLRs. For comparison, Figure~\ref{fig:f13} shows the enhancement/weakening of the ionizing flux received by individual clouds relative to the intrinsic one for the cases with BLR clouds corotating and counterrotating with the BBH. As seen from this figure, the pattern for ionizing flux enhancement/weakening on the $XY$ plane for the BLR clouds in the system with counterrotating BLR clouds (bottom panels) is not only delayed from that for the co-rotating BLR clouds system by phase $\pi$, but also the enhanced and weakened regions in the former case are the reverse of the latter case after correcting the phase delay of $\pi$. 

It might be interesting to consider a case where a fraction of BLR clouds corotate and the rest counterrotate with the BBH system; however, the possibility for such a case may be small because of cloud collisions. The typical collision rate of  a counterrotating cloud with corotating clouds for such a BLR configuration may be roughly estimated as $\sim N_{\rm tot} \Sigma \Delta v/V \sim C \Delta v/ R_{\rm BLR} $ , where $ N_{\rm tot} $ is the total number of clouds in the BLR, $\Sigma$ the mean cross section for the clouds,  $\Delta v$ is the relative velocity of a counterrotating cloud to  corotating clouds if the orbits of BLR clouds are significantly eccentric, and $V$ and $C\sim 0.1$ are the volume and covering factor of BLRs \citep[cf., see][]{2017ApJ...845...88Z}. 
For the BBH-DB-hi model as listed in Table~\ref{tbl:t1}, $\Delta v$ can be up to $\sim 2\times10^4$ \,km/s (on the order of the orbital velocity) and $R_{\rm BLR} \sim 0.06$ \,pc , the collision timescale is  then $\sim 30$\,yr, about one to two times of the orbital period for a cloud along with the BBH systems at a distance $\sim R_{\rm BLR}$. 
Such a BLR configuration with a fraction of clouds corotating while others counterrotating with the central BBH system cannot be stable for a time period longer than a few orbital periods. We note here that the above rough estimate is correct only when those BLR clouds are on substantial eccentric orbits, in which cross-orbits occur frequently. If all BLR clouds are all roughly on circular orbits, then the collision timescale would be much longer. Nevertheless, if a fraction of BLR clouds are corotating and the others are counterrotating in reality, the BEL profile changes would be in between those cases shown in the left and right panels of Figures~\ref{fig:f11} and \ref{fig:f12}. 

For the case of BBH-IntDB, as shown in Figure~\ref{fig:f12}, when the intrinsic variability of the secondary BH dominates the continuum flux variation, the blue- and red-shifted BLR clouds at the same radius are almost photoionized at the same level, with the only fluctuation caused by the position variation of the secondary BH, hence the asymmetric feature caused by the DB effect as shown in Figure~\ref{fig:f11} is weakened. Once the intrinsic variation of the secondary BH contributes $100\%$ variation, the asymmetric characteristics of the profile may be negligible, and thus may not be able to provide useful information on the fraction of corotating and counterrotating clouds only by spectroscopic observations.

\section{Discussions}
\label{sec:discussion}

We have investigated the variations of BELs resulting from those BBH-DB, BBH-Int, and BH-Int systems listed in Table~\ref{tbl:t1}, and further made comparisons between them. In our above calculations, we set the period for all BBH systems to be $2$\,yr, for simplicity. However, the periods of those periodic quasars found in \citet{2015MNRAS.453.1562G} and \citet{2016MNRAS.463.2145C} cover a large range, and the BBH orbital period may be different from the continuum variation period. Below, we briefly discuss the effects on BEL variations of the different choices of $T_{\rm orb}$, and intrinsic variation period ($T_{\rm var}$), with respect to the BLR size.

For the BH-Int model with sinusoidal continuum variation, the BEL variations depend on the BLR size relative to the light-crossing distance over the continuum variation period ($cT_{\rm var}$). If the BLR size is much larger than $cT_{\rm var}$, then the BEL variation is insignificant because the response of BELs to the sinusoidal continuum variation is averaged. In contrast, if the BLR size is much smaller than $cT_{\rm var}$, then the BEL flux variation directly follows the continuum variation. If the BLR size is comparable to $cT_{\rm var}$, then the responses of the BEL photons at different velocities to the continuum variation are different due to the time-delay effect. For the BBH models, the BLR size relative to $cT_{\rm var}$ has a similar effect on the BEL variations. In addition, the periodic change of the ionizing source position in the BBH model also has a significant impact on the BEL profile variations when $R_{\rm RLR}$ is not larger than $a_{\rm BBH}$ by more than an order of magnitude as shown in Section~\ref{subsec:analyse}. In the BBH-DB model ($T_{\rm var}=T_{\rm orb}$), the BEL profile variations are determined by both the DB effect and position change of the ionizing source. The former depends on the velocity of the ionizing source or the compactness of the BBH orbit and the inclination angle, while the latter depends on the ratio $R_{\rm RLR}/a_{\rm BBH}$ (Section~\ref{subsec:analyse}).

In the BBH-IntDB model, $T_{\rm var}$ is not necessarily the same as $T_{\rm orb}$ although we set $ T_{\rm var} = T_{\rm orb}$ in Section~\ref{sec:modelset}. Nevertheless, the effects that arise from the intrinsic variation are significant on the BEL profiles, as long as  $T_{\rm var}$ is comparable with the BLR size.

For those BBH systems investigated in Sections~\ref{sec:modelset} and \ref{sec:results} (Tab.~\ref{tbl:t1}), $T_{\rm orb}$ is fixed at $2$\,yr. However, $T_{\rm orb}$ can be quite different for different BBH systems. If $T_{\rm orb}$ in the BBH-IntDB model is set to much larger than $2$\,yr, then the DB effect is much less significant due to the smaller velocity of the secondary BH. In such a case, the continuum variation may be dominated by the intrinsic variation from the central source. However, the effect on the BEL variation due to the position change of the secondary BH (if it is the central source) is more significant because the ratio of the BBH orbit size to the BLR size is larger. If $T_{\rm orb}$ in the BBH-IntDB model is set to much smaller than $2$\,yr, then the DB effect is more significant but the effect due to the position change of the secondary BH becomes smaller. If the BLR crossing time is much larger than $T_{\rm orb}$, the response of BELs to the periodic continuum variation may be smoothed. 
We do not intend to show examples for the choices of different $T_{\rm orb}$ here because it can be qualitatively understood as the combination of the effects due to DB, position change, and intrinsic variation of the central source, as already discussed in Section~\ref{sec:results} by case studies. 

In our calculations above, we also set a specific spectral index for the continuum $\alpha=-2$ when considering the DB effect. In principle, $\alpha$ at different bands can be quite different from what we adopted, i.e., $\alpha \sim -2$ and $1.1$ in the far-ultraviolet (FUV) and optical bands, respectively, as suggested in \citet{2015Natur.525..351D} for PG1302-102. A different choice of $\alpha$ does affect the significance of the DB effect and thus affects the relative significance of different variation mechanisms on the BEL line variation. However, the patterns of the BEL profile variation due to a different continuum variation mechanism are still more or less the same.  

\section{Conclusions}
\label{sec:conclusion}

In this paper, we investigate the response of the BELs to the continuum variation of optical/UV periodic quasars, under the BBH scenario(s) to interpret the periodicity. We mainly consider two types of models for the periodic continuum variations: one where the variation is mainly caused by the DB-modulated continuum radiation from a disk around the secondary black hole (BBH-DB), and the other where the variation is mainly due to the intrinsic accretion rate variation of the disk around the secondary black hole, though it is also Doppler boosted (BBH-IntDB). We also consider single black hole systems with similar periodic continuum variation caused by intrinsic rate variations (BH-Int), for comparison. For the BLRs of those BBH systems, we adopt simple circumbinary BLR models with standard $\Gamma$ distribution of BLR clouds (emissivity), which is compatible with the BLR size estimates. Our main results on the BELs emitted from those BBH systems and their variations are summarized as follows.

For the BBH systems, the variations of the BEL profiles emitted from the circumbinary BLR are asymmetric (except the extremely face-on-viewed cases), in contrast to that from the BH-Int systems, in which the variations of BEL profiles are symmetric with the blue and red wings enhanced or weakened simultaneously after a delayed response to the central ionization source. With the increasing contribution of the DB effect to the flux variation, the asymmetry of the double-peaked BELs viewed at close to edge-on inclination angles becomes more significant. If the size of the BLR is not dramatically larger than the BBH separation (e.g., $R_{\rm BLR}>10 a_{\rm BBH}$), the position change of the active secondary black hole can also contribute to the asymmetry and variation of the BEL profiles significantly. 

The BBH systems with different variability mechanisms show different behaviors of the BEL profile variations. For the BBH-DB systems, the blue and red wings of the BEL profiles are enhanced and weakened periodically but with a phase difference of roughly $\pi$. For the BBH-IntDB systems, because the variability is dominated by the intrinsic accretion rate variation of the secondary black hole, different from the DB-dominant case, the profile variations of double-peaked BELs show features like the combination of those resulting from the BH-Int and BBH-DB scenarios, i.e., the major part of the variation of a BEL resulting from BBH-IntDB systems is due to the intrinsic variation, with the blue and red parts of the BEL almost simultaneously enhanced or weakened and changing periodically, and a smaller part of the variation is due to the DB effect, with the blue and red parts of the BEL enhanced/weakened periodically but with a phase difference of $\pi$.

For BBH systems, the BEL profiles and their variations also depend on the opening angle of BLR and the fraction of co- and counterrotating clouds in the BLR. For the BBH-DB systems, the larger the opening angle of the BLR, the weaker the effect of continuum variation on the BEL variation. The main reason is that the ionizing continuum emission from the disk around the secondary black hole is significantly enhanced only for those clouds in the region around the direction of movement of the secondary black hole. Furthermore, as a consequence of the combined DB and time-delay effects, if all the clouds are corotating with the binary system, the blue wings fluctuate more strongly than the red wings. Otherwise, the red wings change more strongly than the blue wings if all the BLR clouds are counterrotating. The asymmetry becomes much weaker when the fraction of the corotating clouds and counterrotating clouds are comparable. For those cases with continuum variations caused by the intrinsic rate changes of the disk accretion on the central black hole, the line profile variations are independent of the opening angles and different compositions of co- and counterrotating particles.

The different response patterns of BELs to the continuum variations for different continuum variation models (BBH-DB, BBH-IntDB, BH-Int) may provide an important tool to test or cross-check the BBH scenario(s) for interpreting the optical/UV periodic quasars. Future long-term spectroscopic monitoring and reverberation mapping studies of those periodic quasars may offer a robust way to identify BBHs. 

\acknowledgments{
This work is partly supported by the National Key R\&D Program of China (grant Nos. 2020YFC2201400, 2020SKA0120102, and 2016YFA0400704), the National Natural Science Foundation of China (grant Nos. 11690024, 11873056, and 11991052), the Strategic Priority Program of the Chinese Academy of Sciences (grant No. XDB 23040100), and the Beijing Municipal Natural Science Foundation (grant No. 1204038).}

\end{document}